\documentclass[twocolumn, twocolappendix]{aastex63}

\usepackage{amsmath,amstext}
\usepackage{amssymb}
\usepackage{graphicx}
\usepackage[figure,figure*]{hypcap}
\usepackage[version=3]{mhchem}
\usepackage{mysymbol}
\usepackage{savesym}
\savesymbol{tablenum}
\usepackage{siunitx}
\restoresymbol{SIX}{tablenum}
\usepackage{threeparttable}
\usepackage{url}
\usepackage{comment}
\usepackage{here}

\newcommand{\II}{$\mathrm{I\hspace{-.1em}I}$}
\newcommand{\renv}{r_\mathrm{env}}
\newcommand{\rc}{r_\mathrm{c}}
\newcommand{\tmb}{T_\mathrm{mb}}
\newcommand{\ta}{T^{\ast}_\mathrm{A}}
\newcommand{\rdisk}{R_\mathrm{disk}}
\newcommand{\Fg}{F_\mathrm{G}}
\newcommand{\Fb}{F_\mathrm{B}}
\newcommand{\Fmp}{F_\mathrm{mp}}
\newcommand{\Fmt}{F_\mathrm{mt}}
\newcommand{\Rcurv}{R_\mathrm{curv}}
\newcommand{\vbreak}{v_\mathrm{break}}
\newcommand{\rbreak}{r_\mathrm{break}}
\accepted{Oct.~26, 2021}
\submitjournal{ApJ}

\shortauthors{Sai et al.}
\graphicspath{{./}{figures/}}

\begin{document}

\title{Which Part of Dense Cores Feeds Material to Protostars?: The Case of L1489 IRS}

\correspondingauthor{Nagayoshi Ohashi}
\email{ohashi@asiaa.sinica.edu.tw}

\author[0000-0003-4361-5577]{Jinshi Sai (Insa Choi)}
\affiliation{Department of Astronomy, Graduate School of Science, The University of Tokyo, 7-3-1 Hongo, Bunkyo-ku, Tokyo 113-0033, Japan}
\affiliation{Academia Sinica Institute of Astronomy and Astrophysics, 11F of Astro-Math Bldg, 1, Sec. 4, Roosevelt Rd, Taipei 10617, Taiwan}

\author[0000-0003-0998-5064]{Nagayoshi Ohashi}
\affiliation{Academia Sinica Institute of Astronomy and Astrophysics, 11F of Astro-Math Bldg, 1, Sec. 4, Roosevelt Rd, Taipei 10617, Taiwan}

\author[0000-0002-3801-8754]{Ana\"{e}lle J. Maury}
\affiliation{AIM, CEA, CNRS, Universit\`{e} Paris-Saclay, Universit\`{e} Paris Diderot, Sorbonne Paris Cit\`{e}, 91191 Gif-sur-Yvette, France}
\affiliation{Harvard--Smithsonian Center for Astrophysics, Cambridge, MA02138, USA}

\author[0000-0003-1104-4554]{S\'{e}bastien Maret}
\affiliation{Univ. Grenoble Alpes, CNRS, IPAG, 38000 Grenoble, France}

\author[0000-0003-1412-893X]{Hsi-Wei Yen}
\affiliation{Academia Sinica Institute of Astronomy and Astrophysics, 11F of Astro-Math Bldg, 1, Sec. 4, Roosevelt Rd, Taipei 10617, Taiwan}

\author[0000-0002-8238-7709]{Yusuke Aso}
\affil{Korea Astronomy and Space Science Institute (KASI), 776 Daedeokdae-ro, Yuseong-gu, Daejeon 34055, Republic of Korea}

\author[0000-0002-5482-5206]{Mathilde Gaudel}
\affiliation{LERMA, Observatoire de Paris, PSL Research University, CNRS, Sorbonne Universit\'{e}, 75014 Paris, France}




\begin{abstract}
We have conducted mapping observations ($\sim \ang[angle-symbol-over-decimal]{;2;}\times\ang[angle-symbol-over-decimal]{;2;}$) of the Class I protostar L1489 IRS using the 7-m array of the Atacama Compact Array (ACA) and the IRAM-30m telescope in the \ce{C^18O} 2--1 emission to investigate the gas kinematics on 1000--10,000 au scales. The \ce{C^18O} emission shows a velocity gradient across the protostar in a direction almost perpendicular to the outflow. The radial profile of the peak velocity was measured from a \ce{C^18O} position-velocity diagram cut along the disk major axis. The measured peak velocity decreases with radius at a radii of $\sim$1400--2900 au, but increases slightly or is almost constant at radii of $r\gtrsim$2900 au. Disk-and-envelope models were compared with the observations to understand the nature of the radial profile of the peak velocity. The measured peak velocities are best explained by a model where the specific angular momentum is constant within a radius of 2900 au but increases with radius outside 2900 au. We calculated the radial profile of the specific angular momentum from the measured peak velocities, and compared it to analytic models of core collapse. The analytic models reproduce well the observed radial profile of the specific angular momentum and suggest that material within a radius of $\sim$4000--6000 au in the initial dense core has accreted to the central protostar. Because dense cores are typically $\sim$10,000--20,000 au in radius, and as L1489 IRS is close to the end of mass accretion phase, our result suggests that only a fraction of a dense core eventually forms a star.
\end{abstract}



\section{Introduction} \label{sec:intro}
Low-mass stars are formed through the gravitational collapse of dense cores, having a typical radius of $\sim$0.05--0.1 pc (or 10,000--20,000 au) and mass of $\sim$0.5$\Msun$ \citep{di-Francesco:2007aa, Andre:2014aa, Konyves:2015aa, Pokhrel:2018aa}. 
Comparison between the initial mass function (IMF) and core mass functions (CMFs) of starless cores suggests that only $\sim$40\% of the mass in each dense core contributes to stellar mass \citep{Andre:2010aa, Andre:2014aa, Konyves:2015aa}. If correct, the similarities of the IMF and CMFs implies that most of the material contained in dense cores do not end up in the resulting star(s). However, observationally, it remains unclear that which part of the dense core is eventually accreted onto the central protostar(s), preventing the understanding of star formation efficiency in individual cores. In this paper, we define a zone within the initial core, wherein material forming protostars reside, as ``stellar mass feeding zone'' (SMFZ) and investigate it.

A way to estimate SMFZs around protostars is to measure the radial dependence of the specific angular momentum; infalling material conserving its specific angular momentum can be traced back to its original location within the initial dense core as long as the distribution of its specific angular momentum within the initial dense core is known.

The radial dependence of the specific angular momentum within dense cores have been statistically studied based on the relation between the core size $R$ and the mean specific angular momentum $J/M$, which is well fitted by a power-law $J/M\propto R^{1.6}$ within a radius range of $\sim$0.03--0.3 pc \citep[][]{Goodman:1993aa, Caselli:2002aa, Tatematsu:2016aa}. On the other hand, specific angular momenta $j$ were measured in several infalling envelopes at radii of 200-2000 au, finding that these envelopes show similar specific angular momenta of $\sim$10$^{-3}\kmps$ pc. This finding was interpreted as a consequence of angular momentum conservation during the infalling process \citep[e.g.,][]{Ohashi:1997ab}. \cite{Ohashi:1997ab} have estimated the typical size scale of the SMFZ to be $\sim$6000 au by comparing a heterogeneous ensemble of measurements of the specific angular momentum on different spatial scales \citep[see also][]{Belloche:2013aa}.

Recent observations at higher spatial resolutions have allowed us to directly measure radial profiles of rotational velocity, i.e., specific angular momentum in individual protostellar systems. It has been reported that the specific angular momentum around protostars is often constant within radii of $\sim$100--1000 au \citep{Lee:2010aa, Murillo:2013aa, Yen:2013aa, Harsono:2014aa, Ohashi:2014aa, Aso:2015aa, Aso:2017ab, Yen:2017aa, Maret:2020aa}.
These results are basically consistent with the finding made by earlier studies described above, although these results also suggested that older systems have larger constant specific angular momenta \citep[][]{Yen:2017aa}. On the other hand, radial profiles of the specific angular momentum were measured at larger radii of 800--10,000 au in three protostellar and prestellar objects, showing that the measured radial profiles follow $j\propto r^{1.8}$ \citep{Pineda:2019aa}. This radial dependence of the specific angular momentum is very similar to the relation between the core size and the mean specific angular momentum of dense cores found by \cite{Goodman:1993aa}.

These previous studies have consistently demonstrated that the radial distributions of the specific angular momentum in dense cores and infalling envelopes are divided into two regimes; $j$-constant regime at inner radii and $j$-increase regime at outer radii. \cite{Gaudel:2020aa}, who measured specific angular momentum distributions around 12 Class 0 sources over a wide range of the radius ($\sim$50--5000 au), detected transitions of the two regimes for first time. After taking average of the measured specific angular momentum profiles, although they scatter from a source to a source in some degree, Gaudel et al.~showed that the averaged profile is divided into two regimes at a radius of $\sim$1600 au and the $j$-increase regime is described by $j\propto r^{1.6}$.

\cite{Gaudel:2020aa} have clearly demonstrated that measurements of the specific angular momentum distribution over a wider range of the radius enable us to find the transition between the two regimes of the specific angular momentum distribution. The specific angular momentum distribution in the $j$-increase regime helps us to constrain the specific angular momentum distributions of the initial dense cores, which are required to estimate SMFZs. \cite{Gaudel:2020aa} have also shown that specific angular momentum distributions seem to be different from a source to a source in some degree, suggesting the importance of the measurements for individual sources.

In this paper, we have examined a SMFZ around the Class I protostar L1489 IRS as a case study. L1489 IRS is relatively isolated in the Taurus molecular cloud \citep[d$\sim$140 pc;][]{Zucker:2019aa} and located close to the western edge of the dense core L1489 \citep{Benson:1989aa, Hogerheijde:2000aa, Motte:2001aa, Wu:2019aa}. The object is considered to be a late Class I protostar according to the bolometric temperature and luminosity of 226 K and 3.5$\Lsun$, respectively \citep{Green:2013aa}. Hence, it is a good target to examine a SMFZ. Observations at (sub)millimeter wavelengths reported a dusty envelope around the protostar, which has a radius of $\sim$2000 au and mass of $\sim$0.02--0.03$\Msun$ \citep{Hogerheijde:2000aa, Motte:2001aa}, and rotational motion of the envelope \citep{Hogerheijde:2001aa, Yen:2013aa}. A bipolar outflow is seen in the north-south direction spanning thousands au scales \citep{Tamura:1991aa, Hogerheijde:1998aa, Yen:2014aa}. \cite{Yen:2014aa} have identified a Keplerian disk with a radius of $\sim$700 au, which is the largest one identified around a Class I protostar so far, and two infalling flows accreting onto the disk surface based on their ALMA 12-m array observations in the \ce{C^18O} 2--1 emission at angular resolutions of $\sim\ang[angle-symbol-over-decimal]{;;1}$. \cite{Sai:2020aa} have re-examined the gas kinematics at an angular resolution about three times higher than that in \cite{Yen:2014aa}, revealing that a Keplerian disk with a radius of $\sim$600 au is surrounded by an infalling envelope with a constant specific angular momentum. The dynamical mass of L1489 IRS was also estimated to be $\sim$1.6$\Msun$ \citep{Yen:2014aa, Sai:2020aa}, which is relatively larger than other protostars whose dynamical masses were estimated \citep{Murillo:2013aa, Harsono:2014aa, Chou:2014aa, Lee:2014aa, Aso:2015aa, Aso:2017ab}. The systemic velocity of L1489 IRS was estimated from the Keplerian rotation to be 7.22$\kmps$ \citep{Sai:2020aa}. We have adopted $\vlsr=7.22 \kmps$ for the systemic velocity of L1489 IRS in this paper.

The outline of this paper is as follows. Observations and data reduction are summarized in Section \ref{sec:obs}. Observational results are presented in Section \ref{sec:results}. In Section \ref{sec:analysis}, velocity structures of the \ce{C^18O} 2--1 emission are examined in detail by measuring the peak velocity as a function of radius and comparing it to that measured with kinematic models. We then derive a radial profile of the specific angular momentum of the envelope, discuss its nature, and examine the SMFZ around L1489 IRS in Section  \ref{sec:discussion}. We have also discussed infalling velocity that is slower than the freefall velocity expected from the stellar mass of L1489 IRS. Finally, all results and discussions are summarized in Section \ref{sec:summary}
%
\section{Observations and Data Reduction}\label{sec:obs}
\subsection{IRAM-30m Observations}\label{subsec:obs_30m}
We carried out observations using the IRAM-30m telescope from April 9 to April 14, 2019. Two molecular lines, \ce{C^18O} $J=2$--1 and \ce{N2H+} $J=1$--0, were observed with the heterodyne receivers Eight MIxer Receivers (EMIR) E230 and E090, at the 1.3 mm and 3 mm atmospheric windows, respectively. The rest frequencies are 219.560358 GHz for \ce{C^18O} 2--1 and 93.173770 GHz for \ce{N2H+} 1--0. The \ce{N2H+} 1--0 line is multiplet with seven hyperfine structure (HFS) components \citep[see e.g.,][]{Caselli:1995aa}. Here we provide the rest frequency of the brightest HFS component \ce{N2H+} $J_{F_1,F}=1_{2,3}$--$0_{1,2}$. The VErsatile SPectrometer Arrays (VESPA) backend was connected to the E230 and E090 receivers, providing spectral resolutions of 6.5 kHz ($0.0089 \kmps$) for \ce{C^18O} 2--1 and 20 kHz ($0.063 \kmps$) for \ce{N2H+} 1--0. The velocity resolution for \ce{C^18O} 2--1 was smoothed to $0.17 \kmps$ to increase the signal-to-noise ratio (SNR) and to be combined with ACA data. The observations were carried out in the on-the-fly mapping mode with position-switching, with a reference position located at $\Delta\alpha = -10\arcmin$, $\Delta\delta = -10\arcmin$ with respect to the map center of $\alpha (\mathrm{J}2000)=4^\mathrm{h}4^\mathrm{m}43^\mathrm{s}.07$, $\delta (\mathrm{J}2000)=+\ang[angle-symbol-over-decimal]{26;18;56.2}$, which is a peak position of the 1.3 mm continuum emission in  previous ALMA observations \citep{Yen:2014aa,Sai:2020aa}. We consider this position as the protostellar position throughout this paper. The maps cover a $\sim \ang[angle-symbol-over-decimal]{;2;}\times \ang[angle-symbol-over-decimal]{;2;}$ region around the protostar. The atmospheric opacity at 225 GHz $\tau_{225}$ was $\sim$0.2 in average and ranged from 0.10 to 0.31. The telescope pointing and focus were corrected using Mars every 1--2 hrs and $\sim$4 hrs, respectively. The obtained data set was reduced with the GILDAS software package\footnote{\url{http://www.iram.fr/IRAMFR/GILDAS}}. The antenna temperature $\ta$ was converted to the main beam temperature $\tmb$ using the values of $B_\mathrm{eff}=0.60$ and $F_\mathrm{eff}=0.92$ for \ce{C^18O} 2--1 and $B_\mathrm{eff}=0.81$ and $F_\mathrm{eff}=0.95$ for \ce{N2H+} 1--0, where $B_\mathrm{eff}$ is the main beam efficiency, $F_\mathrm{eff}$ is the forward efficiency, and $\tmb = \ta F_\mathrm{eff}/B_\mathrm{eff}$. The half-power beam width (HPBW) is $\ang[angle-symbol-over-decimal]{;;12}$ for the \ce{C^18O} 2--1 line and $\ang[angle-symbol-over-decimal]{;;28}$ for the \ce{N2H+} 1--0 line. The rms noise levels of the \ce{C^18O} 2--1 and \ce{N2H+} 1--0 images are 110 mK and 42 mK, respectively in $\tmb$. We mainly focus on the \ce{C^18O} emission in this paper. The map of the \ce{N_2H^+} emission is presented in Appendix \ref{sec:app_n2hp}.

We have also mapped a $\ang[angle-symbol-over-decimal]{;10;}$ square region around the protostar in the 1.3 mm continuum with the millimeter camera NIKA2 \citep[][]{Adam:2018aa, Perotto:2020aa} on February 4, 2019, in the pool observations. The observing time was $\sim$0.4 hours and $\tau_{225}$ ranged from 0.24 to 0.31. The obtained data were reduced in the pipeline for NIKA2 data reduction. We then smoothed the obtained map to increase SNR. The final HPBW is $\ang[angle-symbol-over-decimal]{;;20}$, and rms measured at the off-point in the map is 5 $\mjpbm$.
%
\subsection{ACA 7-m array Mosaic Observations}\label{subsec:obs_aca}
Observations using the 7-m array of the Atacama Compact Array (ACA) have been conducted on five nights during December 1--17, 2019, in the ALMA Cycle 7 phase with ten antennae. The baseline length ranged from 8.9--45.0 m, and the minimum baseline provided sensitivity to structures extending to the $\ang[angle-symbol-over-decimal]{;;31}$ scale at a level of 10\% \citep{Wilner:1994aa}. The spectral setup of the observations consisted of five spectral windows for the 1.3 mm continuum and four molecular lines of the \ce{C^18O} 2--1, \ce{^13CO} 2--1 (220.398684 GHz), \ce{^12CO} 2--1 (230.538000 GHz) and \ce{N_2D^+} 3--2 (231.321828 GHz). The spectral windows for the \ce{C^18O} and \ce{^13CO} 2--1 lines, placed within individual basebands, had a bandwidth of 120.0 MHz and a spectral resolution of 61.0 kHz. The spectral windows for the \ce{^12CO} 2--1 and \ce{N_2D^+} 3--2 lines, placed within a single baseband, had a bandwith of 62.5 MHz and a spectral resolution of 61.0 kHz. The spectral window for the 1.3 mm continuum with a bandwidth of 2 GHz was placed within the other baseband. In this paper, we focus on the \ce{C^18O} 2--1 line. The elevation of the target source was $\sim\ang{40}$ in average. The dataset was calibrated in the pipeline with the Common Astronomy Software Applications package \citep[CASA,][]{McMullin:2007aa} version 5.6.1 and its pipeline version 42866M (Pipeline-CASA56-P1-B). The calibration sources were the quasar J0426+2327 for the phase calibration and the quasar J0423-0120 and J0725-0054 for the bandpass and flux calibration. Mosaic observations consisted of 27 fields and covered a $\sim\ang[angle-symbol-over-decimal]{;2;} \times \ang[angle-symbol-over-decimal]{;2;}$ region. The on-source time per pointing was $\sim$2 minutes.

The \ce{C^18O} 2--1 map was produced with CASA 5.6.1 using the \textit{tclean} task. The frequency channels were smoothed to achieve a higher signal-to-noise ratio (SNR) and the resultant velocity resolution was set at $0.17 \kmps$. We adopted Briggs weighting with a robust value of 0.5, the Hogbom deconvolver, and CLEAN masks interactively drawn. The synthesized beam size is  $\sim\ang[angle-symbol-over-decimal]{;;7.7} \times \ang[angle-symbol-over-decimal]{;;6.4}\ (-\ang{85})$ and the rms noise level is $\sim$83$\mjpbm$ ($\sim$43 mK).
%
\subsection{Combination of the ACA and IRAM-30m Data \label{subsec:obs_comb}}
The ACA 7-m array and IRAM-30m images of the \ce{C^18O} emission, which covered similar regions ($\sim\ang[angle-symbol-over-decimal]{;2;} \times \ang[angle-symbol-over-decimal]{;2;}$), were combined for further analysis. Before the images were combined, consistency of the flux scales of the ACA 7-m array and IRAM-30m data was confirmed by comparing the total flux of the two data sets in both the image and uv domain. For this comparison of the flux scales, pseudo-visibility of the IRAM-30m data was produced through the Fourier transform of the IRAM-30m image with the task \textit{uvshort} in CLASS of the GILDAS software.

The two images were combined with the \textit{feather} task in CASA 5.6.1. This task combines interferometer and single-dish images in the $uv$ domain through Fourier transform, weighting them by the spatial frequency response of the images\footnote{See \url{https://casa.nrao.edu/casadocs/casa-5.4.1/image-combination/feather} for more detail.}. The detailed processes of combining images are as follows: (1) the IRAM-30m and ACA 7-m array maps were trimmed to exclude masked and noisy regions at the map edge, (2) the IRAM-30m map was multiplied by the primary beam response of the ACA 7-m mosaic, (3) the primary-beam modified IRAM-30m map and the ACA 7-m array map were combined with the feather task, and (4) the combined map was corrected by the primary beam response of the ACA 7-m mosaic. The synthesized beam size and rms of the combined image are $\ang[angle-symbol-over-decimal]{;;7.7} \times \ang[angle-symbol-over-decimal]{;;6.4}\ (-\ang{85})$ and $\sim$0.14$\jypbm$ ($\sim$70 mK), respectively. We used the combined image for all the analysis of the \ce{C^18O} emission in the following sections.
%
\section{Results}\label{sec:results}
\subsection{Overall Structures of Emissions}
\begin{figure*}
\centering
\includegraphics[width=2\columnwidth]{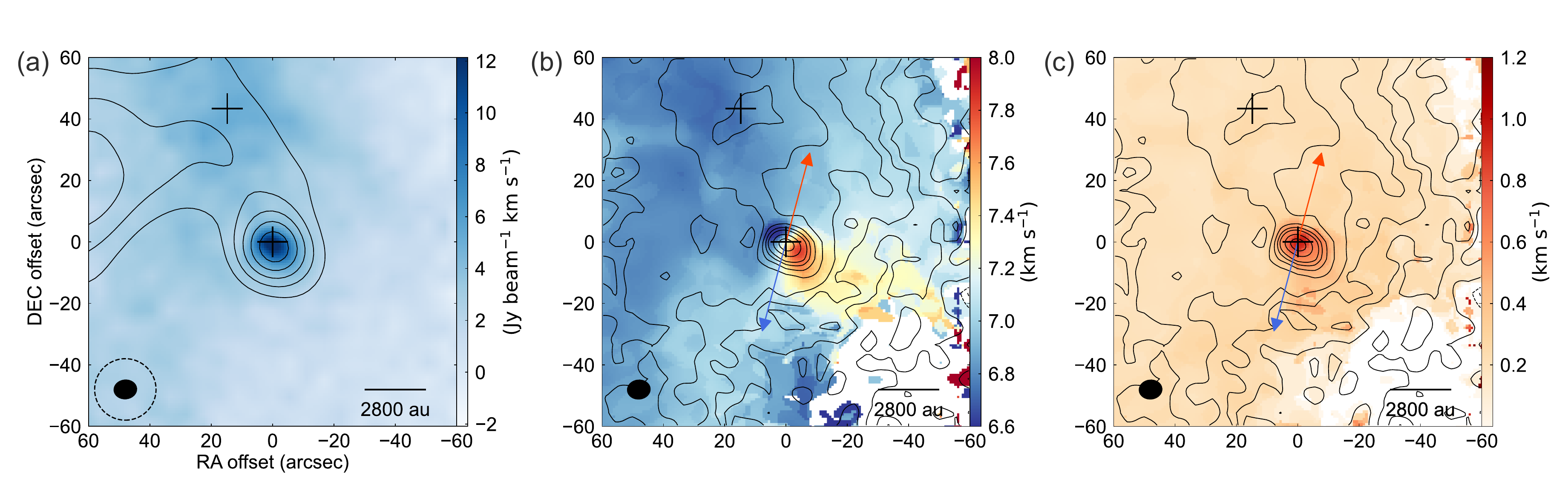}
\caption{(a) The moment 0 map of the \ce{C^18O} 2--1 emission (background color) and the 1.3 mm NIKA2 continuum map (contours). Contour levels start at 3$\sigma$ and increase by steps of 2$\sigma$, where $1\sigma$ is $5\mjpbm$. The central and northern crosses denote the protostellar and \ce{C^18O} second-peak positions, respectively. The filled ellipse and the dashed circle in the bottom-left corner show the beam size of the \ce{C^18O} and continuum maps, respectively. (b) The moment I map of the \ce{C^18O} 2--1 emission (background color) overlaid with the moment 0 map of the \ce{C^18O} 2--1 emission (contours). Contour levels start at 5$\sigma$ and increase by steps of 5$\sigma$ between 5$\sigma$ and 20$\sigma$, steps of 10$\sigma$ between 20$\sigma$ and 60$\sigma$, and steps of 20$\sigma$ between 60$\sigma$ and 100$\sigma$, where $1\sigma$ is $0.12\jypbm\kmps$. Blue and red arrows denote the direction of the blueshifted and redshifted components of the outflow, respectively \citep{Hogerheijde:1998aa}. The filled ellipse in the bottom-left corner shows the beam size of the maps. (c) The same as (b) but the moment \II~map of the \ce{C^18O} 2--1 emission shown in background color. \label{fig:mom_c18o}}
\end{figure*}
Figure \ref{fig:mom_c18o}(a) shows the 1.3 mm NIKA2 continuum map and the velocity-integrated intensity (moment 0) map of the \ce{C^18O} emission. The moment 0 map was calculated by integrating the \ce{C^18O} emission over a velocity range of 5.09--9.34$\kmps$, within which the emission was detected at least above 3$\sigma$ level at the protostellar position. The 1.3 mm continuum emission shows a compact structure with a radius of $\sim\ang[angle-symbol-over-decimal]{;;15}$ ($\sim$2100 au), which is resolved with the smoothed NIKA2 beam, at the protostellar position. This traces a dense, dusty envelope around the protostar. The continuum emission is extended to the northeast from the center and shows the second peak at the east side of the protostar. This second peak originates within a starless core near the protostar, as reported in previous observations \citep{Hogerheijde:2000aa,Motte:2001aa,Wu:2019aa}. The \ce{C^18O} emission exhibits a peak at the protostellar position and it thus likely traces an envelope associated with the protostar. The second peak appears at a position $\sim\ang[angle-symbol-over-decimal]{;;50}$ northeast from the protostar, where the 1.3 mm continuum emission does not present a peak.
%
\subsection{Velocity Structure of the \ce{C^18O} 2--1 Emission}
\begin{figure*}[ht!]
\includegraphics[viewport=40 70 782 525, width=2\columnwidth]{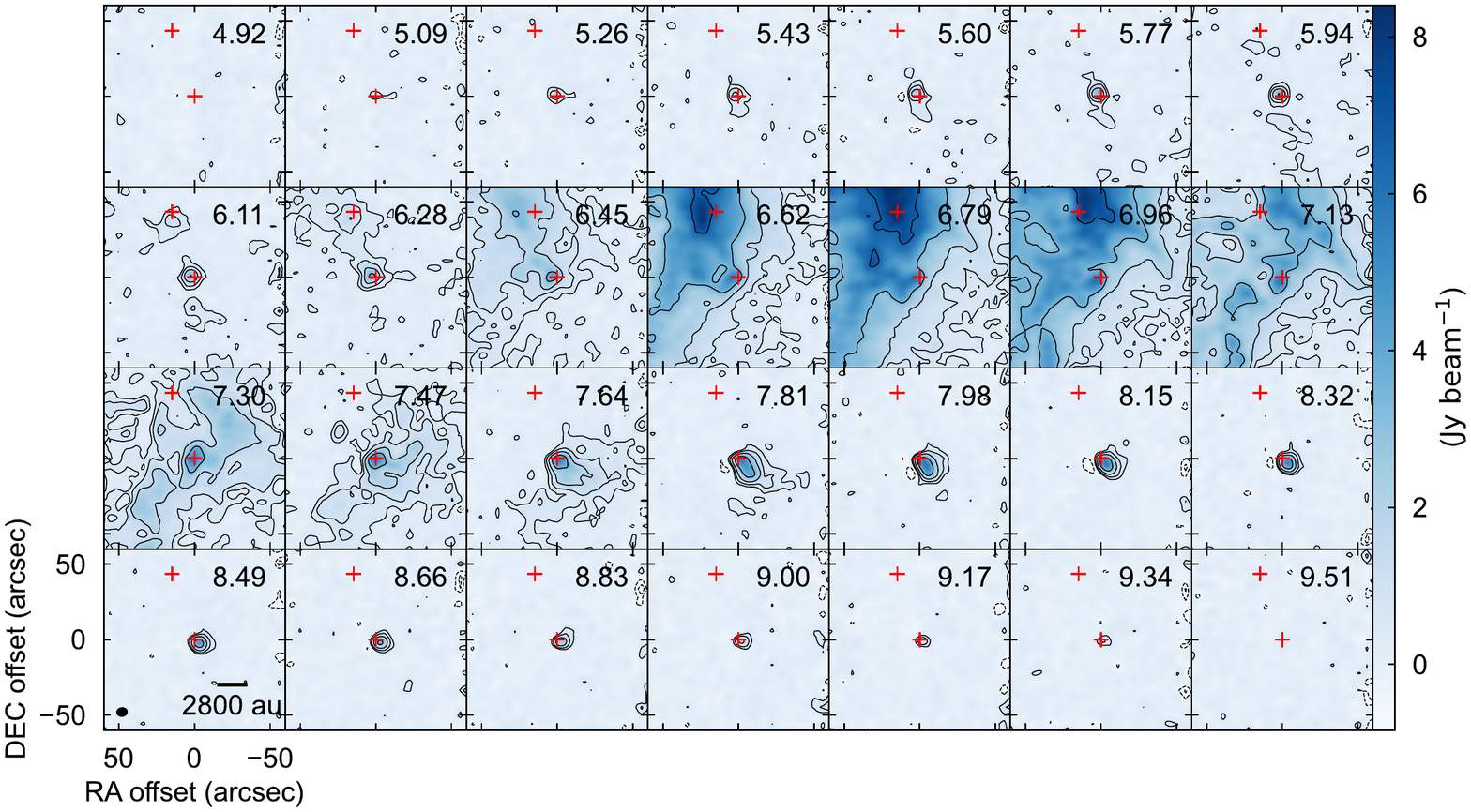}
\caption{Velocity channel maps of the \ce{C^18O} 2--1 emission. Contour levels are 3, 6, 12, 24, and 48 $\times \sigma$, where $1\sigma=0.14\jypbm$. The labels in the top-right corner show the LSR velocity of each channel in$\kmps$. The central and northern crosses denote the protostellar position and the \ce{C^18O} second-peak position measured in the moment 0 map, respectively. The black-filled ellipse in the bottom-left corner denotes the beam size. \label{fig:chan_c18o}}
\end{figure*}
Figure \ref{fig:mom_c18o}(b) provides the velocity-integrated intensity and mean velocity (moment 0 and I) maps of the \ce{C^18O} emission. A velocity gradient is observed from the northeast toward the southwest across the protostellar position. This direction is almost perpendicular to the outflow direction and parallel to that of the velocity gradient due to the rotation of the disk and the envelope seen at radii of $\sim$100--1000 au \citep{Yen:2014aa, Sai:2020aa}, suggesting the velocity gradient of $\gtrsim$1000 au scales is also due to rotational motion. The velocity gradient appears steeper in the vicinity of the protostar and gentler on the outside. To confirm it, the magnitude of the velocity gradient was measured across different scales in the same manner as \cite{Goodman:1993aa} did. We fitted the following function to the mean velocities within a radius of $r_\mathrm{fit}$:
\begin{align}
    \vlsr = v_0 + a \Delta \alpha + b \Delta \delta,
\end{align}
where $v_0$ is the systemic velocity, $\Delta \alpha$ and $\Delta \delta$ are offsets from the protostellar position in right ascension $\alpha$ and declination $\delta$, and $a$ and $b$ are the magnitude of the velocity gradient in the $\alpha$ and $\delta$ directions, respectively. The magnitude and position angle of the velocity gradient are calculated as follows \citep{Maret:2020aa}:
\begin{align}
    G & = \left(a^2  + b^2 \right)^{1/2}/d, \label{eq:mag_grad} \\
    \theta & = \tan^{-1}(a/b), \label{eq:pa_grad}
\end{align}
where $G$ is the magnitude, $\theta$ is the position angle measured from north to east, and $d$ is the distance of the source. Table \ref{tab:vgrad} lists $r_\mathrm{fit}$, $G$, and $\theta$, clearly showing that the magnitude of the velocity gradient is larger in inner regions.
%
\begin{table}[thbp]
    \centering
    \caption{The magnitude and position angle of the velocity gradient measured with the \ce{C^18O} emission}
    \begin{tabular}{ccc}
    \hline
    \hline
    $r_\mathrm{fit}$ & $G$ & $\theta$ \\
     (arcsec) & ($\kmpsppc$) & ($^\circ$) \\
    \hline
    10 & $89 \pm 1$ & $-121.8 \pm 0.7$ \\
    20 & $27.3 \pm 0.2$ & $-125.2 \pm 0.4$ \\
    30 & $14.79 \pm 0.09$ & $-122.0 \pm 0.3$ \\
    40 & $9.58 \pm 0.05$ & $-115.2 \pm 0.3$ \\
    50 & $6.98 \pm 0.03$ & $-110.9 \pm 0.2$ \\
    60 & $5.46 \pm 0.02$ & $-111.7 \pm 0.2$ \\
    \hline
    \end{tabular}
\label{tab:vgrad}
\end{table}

The velocity dispersion (or moment \II) map presented in Figure \ref{fig:mom_c18o}(c) shows that the velocity dispersion decreases rapidly with radius from $\sim$0.7$\kmps$ to $\sim$0.2$\kmps$ within a radius of $\sim\ang[angle-symbol-over-decimal]{;;10}$--$\ang[angle-symbol-over-decimal]{;;20}$ around the protostellar position, while it is almost constant at larger radii.

The velocity structure of the \ce{C^18O} emission is shown in more detail in the velocity channel maps presented in Figure \ref{fig:chan_c18o}. As shown in Figure \ref{fig:mom_c18o}(b), a clear velocity gradient is seen from the northeast to southwest. The emission is compact and within $r\lesssim\ang[angle-symbol-over-decimal]{;;10}$ around the protostar at higher blueshifted ($\vlsr \leq 6.28 \kmps$) and higher redshifted ($\vlsr \geq 8.66 \kmps$) velocities, although there is a weak extension to the northeast at $6.28 \kmps$. The Keplerian rotation was found at these velocities in previous ALMA observations \citep{Yen:2014aa,Sai:2020aa}. Thus, the Keplerian disk component is dominant in the compact emission at these velocities. At lower redshifted velocities ($7.64 \kmps \leq \vlsr \leq 8.49 \kmps$), the emission structures are also relatively compact but with a slight extension of $r\sim\ang[angle-symbol-over-decimal]{;;10}$--$\ang[angle-symbol-over-decimal]{;;30}$ in the southwest direction, even though there is also a weak extension from the northwest to southeast at 7.64$\kmps$. Our previous work has found rotational motion of the infalling envelope around L1489 IRS at these velocities on a scale of $r\lesssim1000$ au $\sim\ang[angle-symbol-over-decimal]{;;7}$ \citep{Sai:2020aa}. These results show that the redshifted emissions trace either the disk or infalling envelope.

On the other hand, the emission structures at lower blueshifted velocities ($6.45 \kmps \leq \vlsr \leq 7.47 \kmps$) are extended across the entire maps. The extended emissions at LSR velocities of 6.62--6.96$\kmps$ appears to be associated with the \ce{C^18O} secondary peak seen in the moment 0 map because they show stronger peaks at the \ce{C^18O} secondary peak position. The molecular emissions associated with L1489 starless core show their peaks at LSR velocities of $\sim$6.6--6.8$\kmps$ \citep[][]{Caselli:2002aa, Wu:2019aa}, suggesting that the extended emission is also a part of the starless core.

%
\begin{figure}[th]
\centering
\includegraphics[width=\columnwidth, viewport=150 0 692 595]{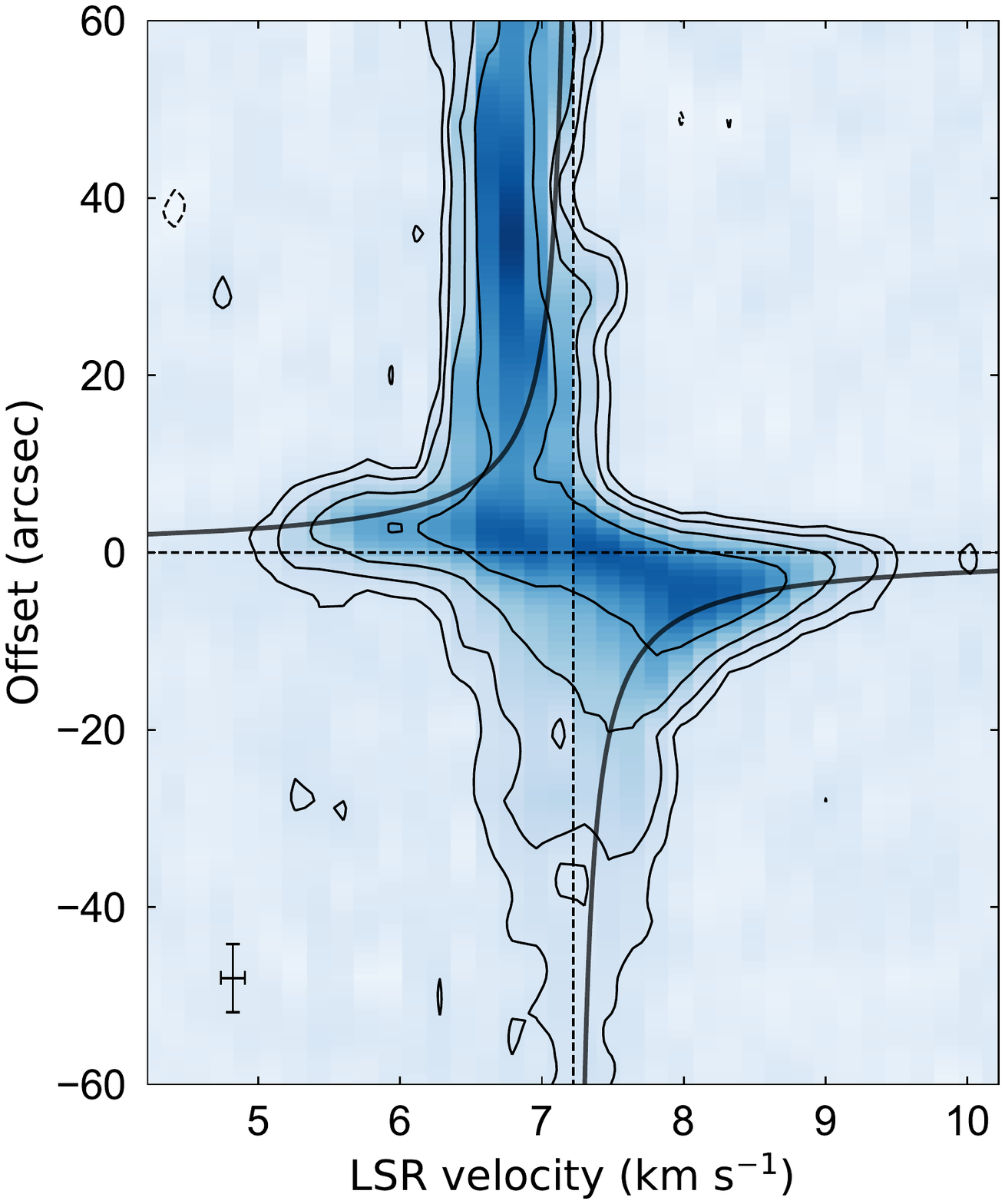}
\caption{Position-velocity (PV) diagram of the \ce{C^18O} 2--1 emission cut along the direction of the disk major axis (P.A.$=\ang{54;;}$). Contour levels are 3, 6, 12, and 24 $\times \sigma$, where $1\sigma=0.14\jypbm$. The black curves indicate the power-law profile $r^{-1}$ of the \ce{C^18O} infalling envelope measured at radii of $\sim$600--1000 au in a previous work \citep{Sai:2020aa}. The vertical and horizontal bars in the bottom-left corner denote the FWHM of the beam major axis and the velocity resolution, respectively. \label{fig:pv_c18o}}
\end{figure}
Figure \ref{fig:pv_c18o} shows a position-velocity (PV) diagram cut along the direction of the disk major axis \citep[P.A.$=\ang{54;;}$;][]{Sai:2020aa} to investigate the velocity gradient of the \ce{C^18O} emission in more detail. It is clear that the velocity of the redshifted emission increases as the position approaches the protostellar position, indicative of differential rotation. The velocity structures of the redshifted emission is mostly consistent with the power-law profile $r^{-1}$ of the \ce{C^18O} infalling envelope measured at radii of $\sim$600--1000 au by \cite{Sai:2020aa}, shown in the black curves in Figure \ref{fig:pv_c18o}, although a part of the emission at $\vlsr > 8.66 \kmps$ arises from the Keplerian disk as was mentioned above. In order to understand how Keplerian and infalling motions contribute to the redshifted emission in the PV diagram, further careful analysis of the velocity structures is required (see Section \ref{subsec:ana_modeling}). 

The velocity structures of the blueshifted emission, on the other hand, appear different from those of the redshifted emission. They are consistent with the $r^{-1}$ profile of the infalling envelope measured in \cite{Sai:2020aa} at velocities of $\vlsr < 6 \kmps$, showing a feature of differential rotation. At velocities of $\vlsr > 6 \kmps$, however, the velocity structures do not follow the $r^{-1}$ profile of the infalling envelope, and the peak velocities are $\sim 0.5 \kmps$ smaller than those expected from the $r^{-1}$ profile. This inconsistency is probably due to the extended emission seen at $\vlsr=$ 6.45--7.47$\kmps$ in the channel maps.
%
\section{Analysis\label{sec:analysis}}
\subsection{Measurement of the Peak Velocity of the $C^{18}O$ 2--1 Emission\label{subsec:ana_rotcurve}}
\begin{figure*}[th]
\centering
\includegraphics[width=1.8\columnwidth]{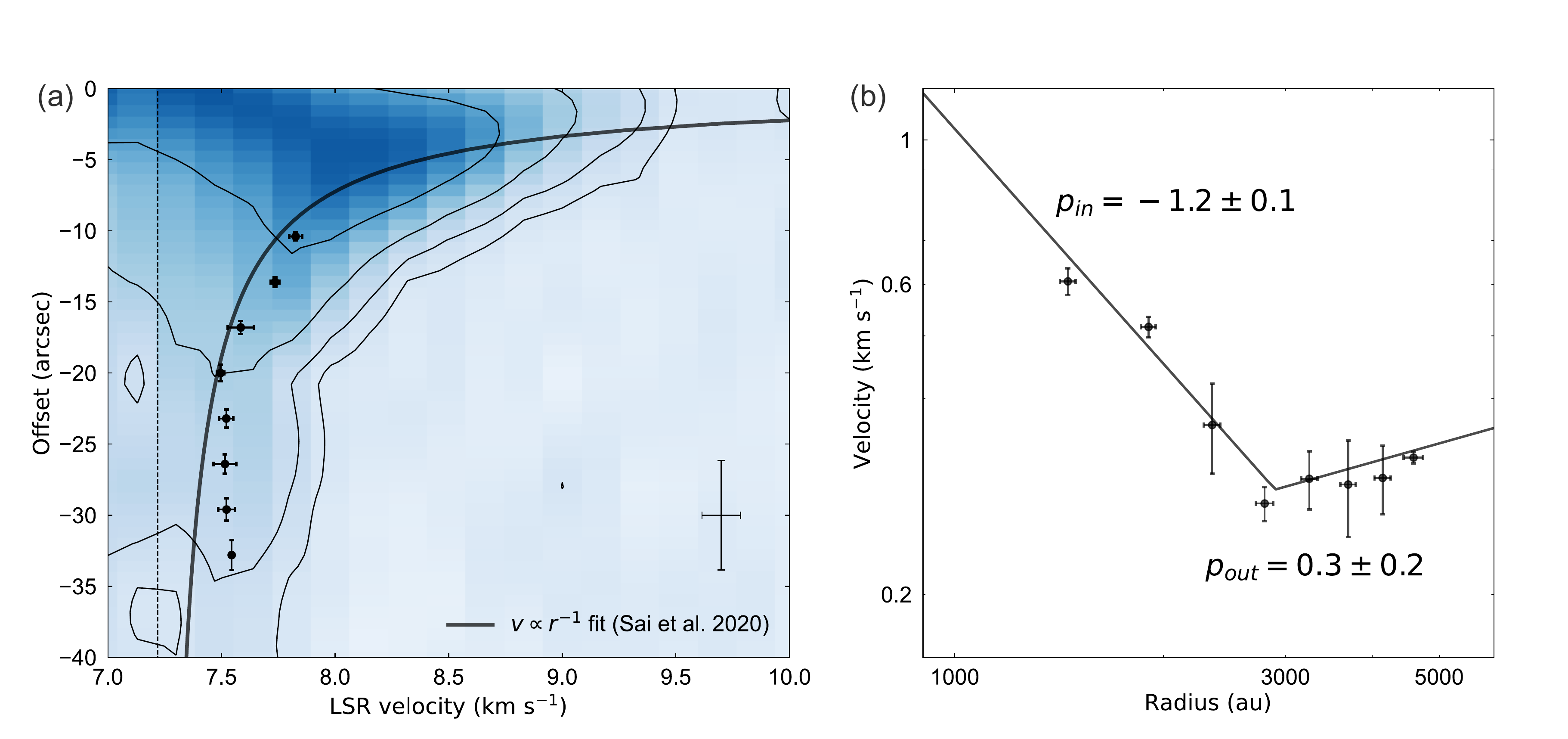}
\caption{(a) Same as Figure \ref{fig:pv_c18o} but with zoomed in offset and velocity ranges of $-\ang[angle-symbol-over-decimal]{;;40}$--$\ang[angle-symbol-over-decimal]{;;0}$ and 7--10 $\kmps$, respectively. Black points represent peak velocities measured through Gaussian fittings to the PV diagram. The vertical and horizontal bars in the bottom-right corner denote the FWHM of the beam major axis and the velocity resolution, respectively. (b) Plots of the data points measured in the PV diagram on a $\log r$-$\log v$ plane. The solid line shows the power-law function with the best-fit parameters.
\label{fig:pvfit}}
\end{figure*}
%
In order to characterize the velocity gradient seen at the redshifted velocity in the PV diagram in Figure \ref{fig:pv_c18o} more quantitatively, the peak velocity was measured as a function of radius in the diagram in the same manner as performed by \cite{Sai:2020aa}. Similar methods were also used in other works \citep{Yen:2013aa,Ohashi:2014aa,Aso:2015aa, Aso:2017ab, Maret:2020aa}. The measurement was only made for the redshifted component because the blueshifted component is highly contaminated by the extended gas associated with the starless core.

A Gaussian function was fitted to the spectrum at each position in the PV diagram to measure peak velocities. The fitting starts from the offset of $-\ang[angle-symbol-over-decimal]{;;10}$ because velocity structures are not spatially well-resolved near the protostar due to the limited angular resolution. The sampling step is $\ang[angle-symbol-over-decimal]{;;3.2}$, which corresponds to half of the beam size. The fitting was performed at positions where the emission was detected at levels above $6\sigma$, resulting in the offset of the outermost point $=-\ang[angle-symbol-over-decimal]{;;35}$. Velocity channels within $\pm$2 channels around a channel, having the maximum intensity, were used for the fitting. This channel selection method traces the peak velocity more accurately than using all the velocity channels when a spectrum is not of a Gaussian shape \citep[see Appendix B in][]{Sai:2020aa}. The uncertainties of the representative data points along position and velocity axes were assumed to be the position accuracy, which is given as angular resolution/(S/N), and fitting error, respectively.
Figure \ref{fig:pvfit}(a) shows the representative data points on the PV diagram.

These data points are plotted on a $\log r$-$\log v$ plane, as presented in Figure \ref{fig:pvfit}(b). The measured peak velocity decreases with radius within $\sim$3000 au, while it slightly increases or is almost constant outside $\sim$3000 au. To characterize the two different behaviors, we fitted the following double power-law function to the data points through $\chi^2$ fitting:
\begin{align}
v_\mathrm{peak} =
	\begin{cases}
	\vbreak \left( \frac{r}{r_\mathrm{break}} \right)^{p_\mathrm{in}}
	 & \left(r \leq r_\mathrm{break} \right) \\
	\vbreak \left( \frac{r}{r_\mathrm{break}} \right)^{p_\mathrm{out}}
	 & \left(r > r_\mathrm{break} \right)
	\end{cases},
\label{eq:dplaw}
\end{align}
where $v_\mathrm{peak} = | \vlsr - \vsys |$. We evaluated uncertainties of the fitting parameters with Monte Carlo method by iterating the fitting procedure 3000 times, changing the offset and velocity values of the data points within their errors. The uncertainties of the fitting parameters were obtained as the standard deviation of the Gaussian distribution fitted to the posterior distributions. In this procedure, uniform and Gaussian prior distributions were assumed for the data errors along the velocity and offset axes, respectively.

A solid line in Figure \ref{fig:pvfit}(b) shows the function with the best-fit parameters of $(\vbreak, \rbreak, p_\mathrm{in}, p_\mathrm{out})=(0.29\pm0.02\kmps, 2900\pm200 \au, -1.2\pm0.1, 0.3\pm0.2)$. The best-fit power-law index at radii smaller than the break radius is close to $-1$. This suggests that the peak velocity mostly trace rotating motion with a constant specific angular momentum, although further careful analysis is required because part of the infalling motion could cause contamination (see the next subsection). On the other hand, the derived power-law index at radii larger than 2900 au is 0.3$\pm0.2$, indicating that the velocity slightly increases or is almost constant with radius.
%
%
\subsection{Disk-and-Envelope Model\label{subsec:ana_modeling}}
In order for us to interpret the origin of the velocity gradient seen in the PV diagram, disk-and-envelope models are constructed and compared with observations in this section.

Although several disk-and-envelope models are examined to find the one explaining observations better in the following discussions, our base model is based on the envelope model proposed by \cite{Ulrich:1976aa} and the disk model discussed in \cite{Sai:2020aa}. The envelope model proposed by \cite{Ulrich:1976aa} is spherical and consists of trajectories of material freefalling from infinite radii and conserving angular momentum. The assumption of the spherical envelope is plausible because the integrated intensity of the \ce{C^18O} emission shows a spherical shape at the protostellar position, as seen in Figure \ref{fig:mom_c18o}. The envelope model is symmetric about the $z$-axis. The velocity and density profiles are given as follows in spherical coordinates:
\begin{align}
    \rho (r, \theta) &= \rho_0 \left(\frac{r}{\rc}\right)^{-1.5} \left(1 + \frac{\cos \theta}{\cos \theta_0} \right)^{-0.5} \nonumber \\
    &~~~~~\times \left(\frac{\cos \theta}{2 \cos\theta_0} + \frac{\rc}{r}\cos^2 \theta_0 \right)^{-1},
    \label{eq:rho_model} \\
    v_r(r, \theta) &= - \left(\frac{GM_\ast}{r} \right)^{0.5} \left(1 + \frac{\cos \theta}{\cos \theta_0} \right)^{0.5},
    \label{eq:vr_model} \\
    v_\theta(r, \theta) & = \left(\frac{GM_\ast}{r} \right)^{0.5} \left(\cos \theta_0 - \cos \theta \right) \nonumber \\
    &~~~~~ \times \left(\frac{\cos\theta_0 + \cos \theta}{\cos \theta_0 \sin^2 \theta} \right)^{0.5},
    \label{eq:vtheta_model}\\
    v_\phi(r, \theta) &= \left(\frac{GM_\ast}{r} \right)^{0.5} \frac{\sin \theta_0}{\sin \theta} \left(1 - \frac{\cos \theta}{\cos \theta_0} \right)^{0.5}.
    \label{eq:vphi_model}
\end{align}
Here $G$ is the gravitational constant, $M_\ast$ is the stellar mass, $\theta_0$ is the initial location of the infalling material in $\theta$ at an infinite radius and $\rc=\frac{j^2}{G M_\ast}$ is the centrifugal radius, where $j$ is the specific angular momentum. The reference density $\rho_0$ is defined as $\rho=\frac{2\rho_0}{3\sqrt{2}}$ at $(r,\theta)=(\rc, 0^\circ)$.

Ulrich's envelope model has five parameters of $M_\ast$, $\rc$, $\renv$, $i$, and $\rho_0$, where $\renv$ is the envelope radius and $i$ is the inclination angle of the system. First, we adopted the following values for the parameters based on observations for a base model. For the protostellar mass, $M_\ast=1.6 \Msun$ derived from the Keplerian rotation of the disk was adopted \citep{Yen:2014aa,Sai:2020aa}. $\rc=\rdisk=600$ au was assumed for $\rc$, as measured by \cite{Sai:2020aa} and $\renv$ was fixed at 5000 au based on the extent of the \ce{C^18O} emission. The inclination angle of the system was assumed to be $\ang{73}$, according to the inclination angle of the disk estimated from the aspect ratio of the dust continuum emission \citep{Sai:2020aa}. Although the inclination angle is uncertain, the model results do not strongly depend on it if it is greater than $\ang{60}$, as is the case with L1489 IRS \citep{Stark:2006aa, Brinch:2007aa,Eisner:2012aa,Sheehan:2017aa}. The volume density $\rho_0$ was initially fixed at $1.4\times 10^{-18}\ \mathrm{g\ cm^{-3}}$, which provides the intensity of \ce{C^18O} almost consistent with the observations, while it was adjusted when our models were modified to explain the observations better.

%
\begin{figure*}[ht]
\centering
\includegraphics[width=2\columnwidth]{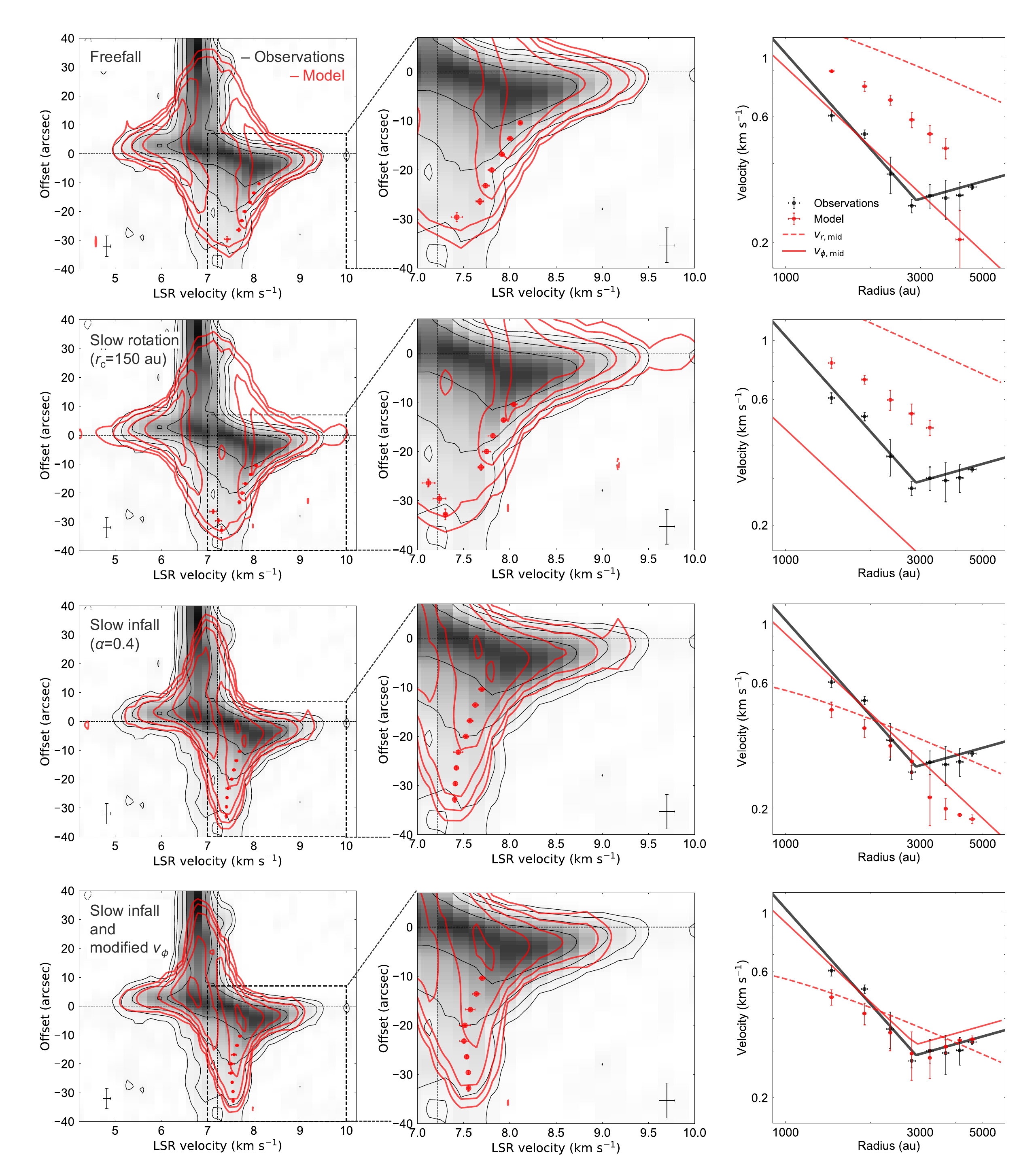}
\caption{(First and second columns) Comparison of PV diagrams cut along the disk major axis between the observations (black contours and grey scale) and models (red contours). Contour levels of both observed and model PV diagrams are the same as those in Figure \ref{fig:pv_c18o}. Red data points are the peak velocities measured with the model PV diagrams. (Third column) Radial profiles of the measured peak velocity. Black and red data points are peak velocities measured with the observed and model PV diagrams, respectively. Black, solid lines represent the best-fit power-law function for the data points from the observations. The red, solid and dashed lines show the azimuthal- and radial-velocity components along the disk mid-plane in the models, i.e., $v_\phi(r, \theta=90º)$ and $v_r(r, \theta=90º)$, respectively, described with Equation (\ref{eq:vr_model}), (\ref{eq:vphi_model}) and (\ref{eq:vphi_jinc}). From the first to the bottom rows, results of the model with freefall, slow rotation, slow infall and modified $v_\phi$ are presented. \label{fig:pv_obsvsmodels}}
\end{figure*}
%
Regarding the disk in our models, the disk model discussed in \cite{Sai:2020aa} was adopted except for its warped and gap structures. The warped and gap structures affect only the morphology at $r<$ 600 au and are thus negligible in comparison with the current observations with an angular resolution of $\sim$1000 au. Note that the same inclination and position angles were assumed for both the disk and the envelope, although they do not necessarily have the same rotational axis as suggested in the past observations \citep{Brinch:2007aa,Sai:2020aa}. All the parameters for the base model are summarized in Table \ref{tab:model_params}. Because the infall velocity is freefall in the base model, the base model is called the disk-and-envelope model with freefall in the following discussions.
%
\begin{table}[thbp]
\centering
\begin{threeparttable}
    \begingroup
    \centering
    \caption{Parameters of the base disk-and-envelope model}
    \begin{tabular*}{\columnwidth}{@{\extracolsep{\fill}}lc}\hline\hline
    Parameter & Value \\\hline
    \textit{Protostellar and global parameters} &  \\
    \quad Protostellar mass ($M_\ast$) & $1.6 \Msun$ \\
    \quad \ce{C^18O} abundance ($X_\mathrm{C^{18}O}$) & $1.7\times10^{-7}$ \\
    \textit{Disk parameters} &  \\
    \quad Disk mass ($M_\mathrm{disk}$) & $0.0071 \Msun$ \\
    \quad Power of surface density profile ($p$) & 0.5 \\
    \quad Disk radius ($\rdisk$) & $ 600 \au$ \\
    \quad Position angle (P.A.) & $\ang{54}$ \\
    \textit{Envelope parameters} &  \\
    \quad Centrifugal radius ($\rc$) & 600 au \\
    \quad Envelope radius ($\renv$) & 5000 au \\
    \quad Inclination angle ($i$) & $\ang{73}$ \\
    \quad Density ($\rho_0$) & 1.4$\times10^{-18}~\mathrm{g\ cm^{-3}}$  \\
    \hline
    \end{tabular*}
    \label{tab:model_params}
	\endgroup
\end{threeparttable}
\end{table}
%
%
%

In order to compare our models with the observations, synthetic observations were carried out based on our models. The intensity scales of the synthetic observations were based on the radiative transfer calculations using RADMC-3D\footnote{\url{http://www.ita.uni-heidelberg.de/~dullemond/software/radmc-3d/}} with the assumption of the LTE condition. The opacity table derived by \cite{Semenov:2003aa}\footnote{\url{https://www2.mpia-hd.mpg.de/home/henning/Dust_opacities/Opacities/opacities.html}} was adopted. The protostellar temperature $\sim$4000 K calculated from 3.5 $L_\mathrm{bol}$ was used for the protostellar SED to calculate the temperature profile within the envelope and the disk. The molecular abundance of \ce{C^18O} to \ce{H2} gas and gas-to-dust mass ratio were assumed to be $1.7\times10^{-7}$ and 100, respectively \citep{Frerking:1982aa}. We added Gaussian noise to the model images and convolved them with a Gaussian beam so that they have the same resolution and SNR as the observations. PV diagrams were generated from the calculated model images and compared to the observed PV diagram.

The first row in Figure \ref{fig:pv_obsvsmodels} shows the comparison between the disk-and-envelope model with freefall and the observations. As discussed in Section \ref{sec:results}, the blueshifted emission in the observations is not likely associated with L1489 IRS. Hence, comparisons were made only for the redshifted components. PV diagrams zoomed in a velocity range of 7 to 10$\kmps$ and an offset range of $-\ang[angle-symbol-over-decimal]{;;40}$ to $\ang[angle-symbol-over-decimal]{;;7}$ are presented in the second column in Figure \ref{fig:pv_obsvsmodels} to compare the redshifted components between the observations and models. As seen in the second panel in the first row, the model and observed PV diagrams are not consistent. The model PV diagram shows intensity peaks at higher velocities than the observations, and is wider along the offset axis than in the observations. In order to make further comparisons, the peak velocity was measured in the model PV diagram in the same way as that used for the observed PV diagram in Section \ref{subsec:ana_rotcurve}. The third column in Figure \ref{fig:pv_obsvsmodels} shows the radial profiles of the measured peak velocity on the $\log r$-$\log v$ plane. The red, solid and dashed lines show the rotational and infalling velocity along the disk mid-plane in the models, i.e., $v_\phi(r, \theta=90º)$ and $v_r(r, \theta=90º)$, respectively, described with Equation (\ref{eq:vr_model}), (\ref{eq:vphi_model}) and (\ref{eq:vphi_jinc}). The third panel in the first row shows that the measured velocity for the freefall model is much higher than that from the observations, suggesting that infalling or rotational velocity in the model is too large to explain the observations.

To examine a case where the rotational velocity is suppressed, we produced a disk-and-envelope model with slow rotation with $\rho_0 = 1.1\times 10^{-17}\ \mathrm{g\ cm^{-3}}$ and $\rc=150$ au, which corresponds to a rotational velocity two times slower than that in the model with freefall. Note that the reference density was also modified because it is defined as a value at $\rc$. The second row in Figure \ref{fig:pv_obsvsmodels} compares the model with slow rotation model and the observations. The first panel shows that the overall shape of the PV diagram of the model with slow rotation is very similar to that of the model with freefall but the former is more axisymmetric about the offset axis. Regardless of the slower rotational velocity, the intensity peaks appear at higher velocities than the observations, as seen in the second panel. The measured peak velocity was also much higher than that of the observations, as shown in the third panel. These results suggest that the peak velocity is almost determined by the infalling velocity in the case where the infalling velocity is freefall.

The infalling velocity in the two models are too high to explain the observations. Although one might consider to adopt a smaller $M_\ast$ to reduce the infalling velocity in the model, $M_\ast$ is well-constrained from Keplerian rotation of the disk. Thus, we introduced a new parameter $\alpha$ as $v_r = \alpha v_{r,\mathrm{ff}}$ to suppress the infalling velocity, maintaining the same stellar mass. Here, $v_{r,\mathrm{ff}}$ is the radial velocity in the case of the freefall expressed by Equation (\ref{eq:vr_model}). Strictly speaking, Ulrich's envelope model does not show self-consistency with the additional parameter $\alpha$; nonetheless, this approach would be still acceptable to observe a tendency of the velocity structures with suppressed infalling velocity. As a case where the infalling velocity was slower than the freefall velocity, a model with $\alpha=0.4$ was produced and compared to the observations. In the second panel, in the third row in Figure \ref{fig:pv_obsvsmodels}, the model with slow infall reproduces the overall shape of the observed PV diagram well. The radial profile of the measured peak velocity, indicated in the third panel, is also consistent with the profile measured from the observations at radii smaller than the break radius of $\sim$2900 au. Although the measured peak velocity is a blend of the rotational velocity and infalling velocity, its absolute values are very similar to the given rotational velocity.

Among the three disk-and-envelope models discussed above, the one with slow infall ($\alpha=0.4$) reproduces the observations best, suggesting that the infall motion in the envelope is slower than the freefall motion. Nevertheless, we should stress that none of them can reproduce the radial profile of the observed peak velocities at radii larger than 2900 au. The nature of the radial profile of the measured peak velocity, including the velocity increase outside the break radius, will be further discussed in the next section.

%
\section{Discussion \label{sec:discussion}}
\subsection{Nature of the Radial Profile of the Measured Peak Velocity \label{subsec:nature_vprof}}
The radial profile of the measured peak velocity shows a break at a radius of $\sim$2900 au, as shown in Section \ref{subsec:ana_rotcurve}. The profile inside the break radius is consistent with a $r^{-1}$ power-law and this is indeed expected if the motions are dominated by rotation in an infalling and rotating envelope with a constant $j$, as was demonstrated using the disk-and-envelope model with slow infall.
One might wonder whether the observed velocity structures could originate from non-axisymmetric infalling flows suggested in a previous work \citep[][]{Yen:2014aa} rather than rotational motion of a more spherical envelope. However, the extent of the suggested infalling flows is only $\sim\ang[angle-symbol-over-decimal]{;;10}$ on the plane-of-sky, while the emission explained by the spherical envelope model is extending up to a radius of $\sim\ang[angle-symbol-over-decimal]{;;20}$ from the protostellar position. The previous ALMA observations in \cite{Yen:2014aa} would have detected only the density-enhanced non-axisymmetric infalling flows within a more spherical envelope due to missing flux. The limited field of view of the previous ALMA observations, whose FWHM of the primary beam is $\sim15''$, could also be a reason that the emission extending up to a radius of $\sim\ang[angle-symbol-over-decimal]{;;20}$ from the protostellar position was not detected in \cite{Yen:2014aa}. The infalling flows reported in the previous work are almost along the north-south direction, which also suggests that they are not likely responsible for the velocity structure in the PV diagram cut along the disk major axis (P.A. $=54^\circ$) in the current work.

The measured peak velocity outside the break radius, on the other hand, increases or is almost flat. A simple explanation of the increase of the measured peak velocity outside the break radius may be caused by a change in the radial dependence of the rotational velocity at the break radius. Such a change is possible if the infalling material carries a greater specific angular momentum at outer radii in the envelope \citep[][]{Yen:2011aa, Takahashi:2016aa}. In order to demonstrate this possibility, the disk-and-envelope model with slow infall was modified to have the following rotational velocity\footnote{Equation (\ref{eq:vphi_jinc}) for $r\le \rbreak$ is another expression of Equation (\ref{eq:vphi_model}), and derived from Equation (10) in \cite{Ulrich:1976aa} and Equation (\ref{eq:vphi_model}).}:
\begin{align}
v_\phi(r, \theta) =
    \begin{cases}
    v_\mathrm{break} \left(
    \frac{r}{\rbreak}
    \right)^{-1} \frac{\sin \theta_0}{\sin \theta} & (r \le \rbreak) \\
    v_\mathrm{break} \left(
    \frac{r}{\rbreak}
    \right)^{p_\mathrm{out}} \frac{\sin \theta_0}{\sin \theta} & (r > \rbreak)
    \end{cases},
    \label{eq:vphi_jinc}
\end{align}
where $v_\mathrm{break} = \sqrt{G M_\ast \rc \sin{\theta_0}^2}/\rbreak$, $\rbreak=2900$ au, and $p_\mathrm{out}=0.3$. The PV diagram of the model with the slow infall and modified $v_\phi$, presented in the fourth row in Figure \ref{fig:pv_obsvsmodels}, is very similar to that obtained with the original model with slow infall (before the modification), shown in the third row, while the radial profile of the peak velocity obtained from the model with slow infall and modified $v_\phi$ reproduces the observed radial profile well even at radii larger than 2900 au. The increase of the measured peak velocity could be caused by a change in the radial dependence of the infalling velocity instead of the rotational velocity. Disk-and-envelope models with such changes, however, fail to explain the observations, as demonstrated in Appendix \ref{subsec:app_vrmodel}.

Note that the break in the profile could also occur if the dominant mechanism making the velocity gradient changes from rotation to another such as infalling flow and turbulence. Although non-axisymmetric infalling flows were suggested in a previous work \citep[][]{Yen:2014aa}, these flows have the extent only $\sim\ang[angle-symbol-over-decimal]{;;10}$ ($\sim$1400 au) and are almost along the north-south direction as was mentioned above. Hence, we examined whether an infalling flow coming from another direction could reproduce the increase of the peak velocity outside the break radius in Appendix \ref{subsec:app_flowmodel}. As shown in Figure \ref{fig:app_model_flow_pv}, the infalling flow model barely reproduces the increase of the peak velocity outside 2900 au, although the direction of the infalling flow has to be very specific, as explained in Appendix \ref{subsec:app_flowmodel}. Another possible motion to explain the increase of the peak velocity would be cloud-scale turbulence characterized by Larson's law \citep[$\Delta v \propto r^{\sim0.5}$;][]{Larson:1981aa,McKee:2007aa}. Cloud-scale turbulence can produce a velocity gradient on sub-parsec scales in projected maps \citep{Burkert:2000aa}, and the velocity deviation is expected to be proportional to $r^{0.5}$. Thus, the increase of the peak velocity outside the break radius could be explained as the tail of the scaling law of the cloud-scale turbulence, as also discussed by \cite{Gaudel:2020aa}. On the other hand, the velocity gradient in L1489 IRS appears systematic and coherent. Table \ref{tab:vgrad} shows that the directions of the velocity gradient have only a small dispersion of $\sim\pm$5$^\circ$ over a spatial scale of $\ang[angle-symbol-over-decimal]{;;10}$--$\ang[angle-symbol-over-decimal]{;;60}$ ($\sim$1400--8400 au). Numerical simulations suggest that the angular momentum axis often varies by $\sim$20--90$^\circ$ with a radius from 1000 to 10,000 au in turbulent dense cores \citep[][]{Joos:2013aa, Matsumoto:2017aa}. The velocity gradient in the projected map is also expected to change the directions with radius for such a turbulent velocity field. Hence, it would be more natural to interpret the coherent velocity structure of L1489 IRS as rotational motion rather than turbulence.

In summary, a simple and probably more natural explanation of the increase of the measured peak velocity outside the break radius is that the radial dependence of the rotational velocity changes at the break radius. An infalling flow or turbulence could also explain the increasing peak velocity if such motions would be more dominant than rotation outside the break radius. In the following sections, we further discuss the simple and more natural case where the radial dependence of the rotational velocity changes at the break radius.

\subsection{Specific Angular Momentum Profile in L1489 IRS \label{subsec:ang_prof}}
\begin{figure}[th]
\centering
\includegraphics[width=\columnwidth]{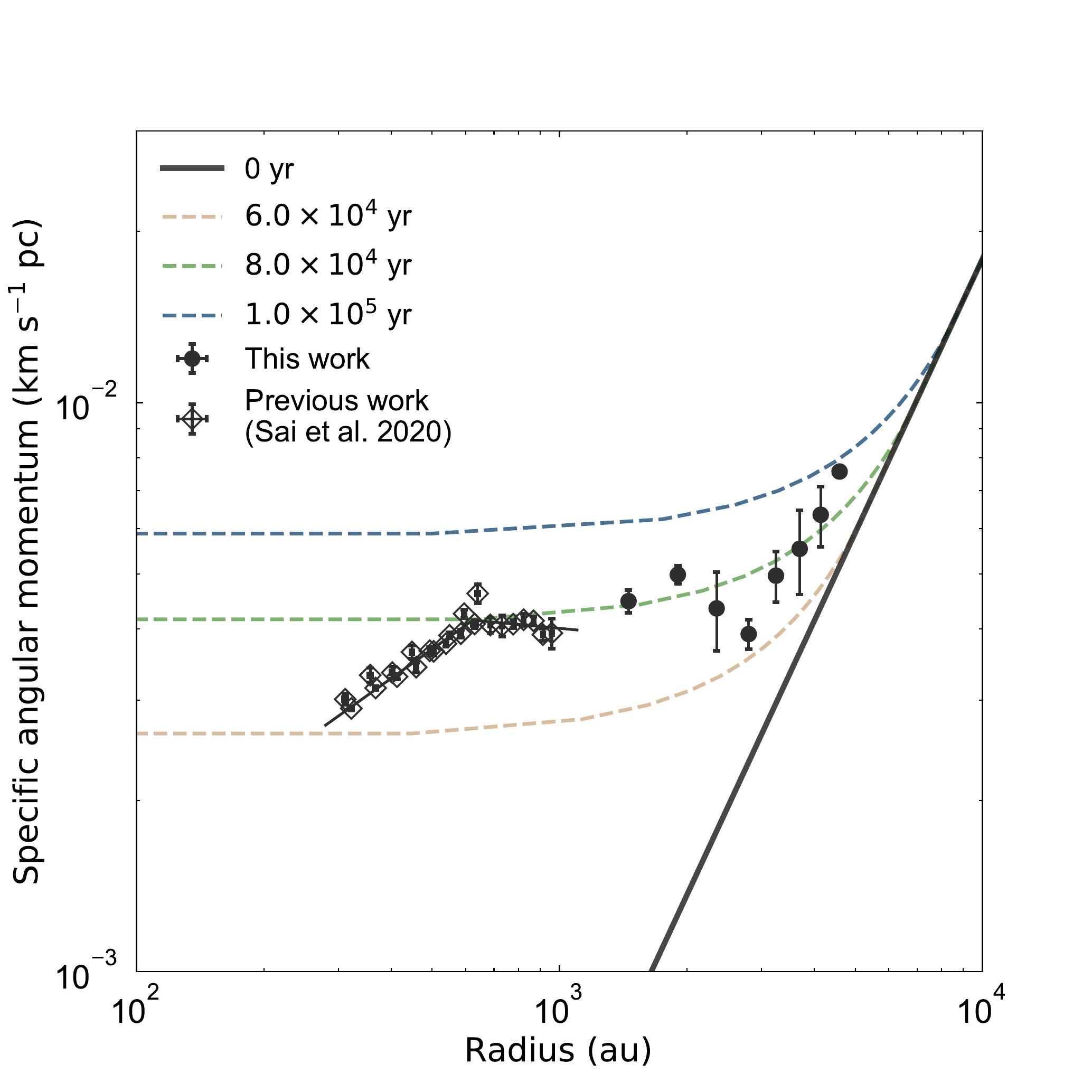}
\caption{Radial profile of the specific angular momentum measured in L1489 IRS (data points) and those at certain time in a calculation of the inside-out collapse model (solid and dashed lines). \label{fig:j_insideout}}
\end{figure}

As the radial profile of the measured peak velocity is well explained by the rotation motion of the envelope, the specific angular momentum can be calculated from the measured peak velocity with the relation of $j = r \times v$ after a correction of the inclination angle of $\ang{73}$. Again, the measured peak velocity is not exactly, but almost, comparable to the rotational velocity, as shown in the comparison to the models. The difference between the measured peak velocity and the expected rotational velocity, with a correction of the inclination angle in the slow-infall model, is about 10\%. The specific angular momentum of the disk and envelope within 1000 au was also derived from the rotation curve measured in the previous work of \cite{Sai:2020aa}. The derived specific angular momentum is plotted as a function of the radius in Figure \ref{fig:j_insideout}. The uncertainty of the derived specific angular momentum due to the uncertainty of the inclination angle between 60--$\ang{90}$ is roughly the size of the symbol in the plot.
Figure \ref{fig:j_insideout} shows that the specific angular momentum profile consists of three different regimes: a Keplerian disk inside $\sim$600 au, a $j$-constant regime at $\sim$600--2900 au, and a $j$-increase regime outside $\sim$2900 au.

A kinematic transition between the $j$-constant and $j$-increase regimes has been reported in Class 0 sources \citep{Gaudel:2020aa}. They calculated apparent specific angular momenta from LSR velocities measured at radii of 50--5000 au for 12 Class 0 protostellar systems. Although the obtained radial dependences of the apparent specific angular momentum were largely scattered from one source to another, the mean profile showed a break at a radius of $\sim$1600 au. The mean apparent specific angular momentum in the $j$-constant regime is $\sim$6$\times 10^{-4}\kmps \pc$. These values are smaller than those in L1489 IRS, $\rbreak$ of $\sim$2900 au and $j$ of $\sim$4$\times 10^{-3}\kmps \pc$. This is consistent with simple analytical calculations of angular momentum transfer from an initial dense core where specific angular momentum is larger at a larger radius \citep[][]{Yen:2011aa, Takahashi:2016aa,Yen:2017aa}.

%
\begin{figure}[th]
\centering
\includegraphics[width=\columnwidth]{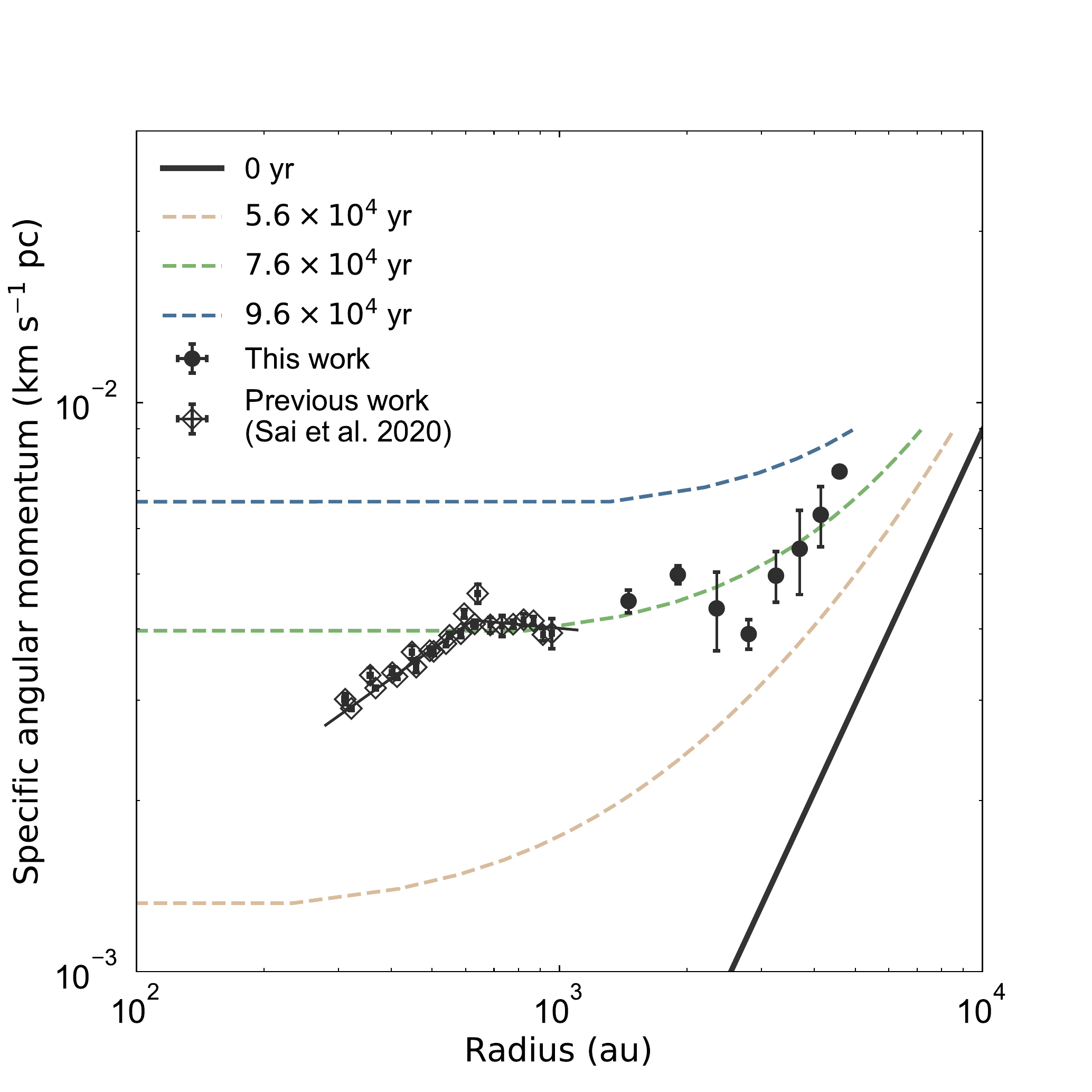}
\caption{Same as Figure \ref{fig:j_insideout} but with the result for the model based on \cite{Takahashi:2016aa}. \label{fig:j_T16}}
\end{figure}

In order to further investigate the nature of the specific angular momentum profile and to discuss the size of the SMFZ, we performed simple, analytical model calculations of the angular momentum transfer from an initial dense core to an infalling envelope and compared the calculated radial profiles of the specific angular momentum to that measured in the observations.

First, we calculate the angular momentum transfer along an equatorial plane of an initial dense core, based on the inside-out collapse of a singular isothermal sphere with a finite angular momentum \citep{Shu:1977aa, Yen:2011aa, Yen:2017aa}. We assume that the specific angular momentum profile in the initial dense core is given by
\begin{align}
    j = j_0 \left(\frac{r}{10^4 \au} \right)^{1.6},
    \label{eq:j_model}
\end{align}
as suggested by previous observations of dense cores \citep{Goodman:1993aa, Caselli:2002aa, Tatematsu:2016aa, Pineda:2019aa, Gaudel:2020aa}. The density distribution of the initial dense core is assumed as follows \citep{Shu:1977aa}:
\begin{align}
    \rho = \frac{\cs^2}{2 \pi G} r^{-2},
    \label{eq:rho_shu}
\end{align}
where $\cs$ is the isothermal sound speed. The enclosed mass including the central object mass is calculated from the density distribution as
\begin{align}
    M(r) = \frac{2 \cs^2}{G} r.
\end{align}
Assuming that infalling material follows freefall, infalling motion of material is calculated from the equation of motion of the gas element:
\begin{align}
    \frac{dv_r}{dt} = - \frac{GM}{r^2}.
    \label{eq:eq_of_motion_grav}
\end{align}
Here $M$ is the enclosed mass within the initial radius $r_\mathrm{ini}$, at which the infalling material is initially located. The lefthand side of the equation can be written as
\begin{align}
    \frac{dv_r}{dt} = \frac{d r}{dt} \frac{d v_r}{dr} = \frac{1}{2} \frac{d v_r^2}{dr}.
\end{align}
By substituting this in Equation (\ref{eq:eq_of_motion_grav}) and integrating, the following equations are obtained:
\begin{align}
   \frac{1}{2}v_r^2 & = \int_{r_\mathrm{ini}}^r - \frac{GM}{r'^2} dr^\prime, \nonumber \\
   \therefore v_r = \frac{dr}{dt} &= - \left\{
   2GM \left( \frac{1}{r} - \frac{1}{r_\mathrm{ini}} \right)
   \right\}^{1/2},\label{eq:vr_int}
\end{align}
and thus,
\begin{align}
dt = - \left\{
   2GM \left( \frac{1}{r} - \frac{1}{r_\mathrm{ini}} \right)
   \right\}^{-1/2} dr. \label{eq:t_int}
\end{align}
Note that $M=M(r_\mathrm{ini})$ is constant on the frame of the gas element and $v_r$ is negative because infalling motion is considered. By replacing $r$ with the normalized radius $x\equiv \frac{r}{r_\mathrm{ini}}$ and integrating Equation (\ref{eq:t_int}), the relation between the radial location and time is obtained.
\begin{align}
    t = - \sqrt{\frac{r_\mathrm{ini}^3}{2GM}} \int_{1}^{x}  \frac{dx'}{\sqrt{x'^{-1} - 1}}.
    \label{eq:t_x_insideout}
\end{align}

In the inside-out collapse model, the expansion wave propagates from the inside to the outside at the isothermal sound speed $c_\mathrm{s}$ and gas located at a certain radius starts to infall once the expansion wave has reached there. Therefore, we set $t=0$ at the time when the innermost part of the core collapses, and a gas element at $r=r_\mathrm{ini}$ starts infalling when the expansion wave reaches $r_\mathrm{ini}$, i.e., at $t= t_\mathrm{wave} = r_\mathrm{ini}/c_\mathrm{s}$. The radial distribution of the specific angular momentum at $t$ is then derived assuming the angular momentum conservation, $j(r(t))=j(r_\mathrm{ini})$. The protostellar mass at $t$ is also estimated from $M(r_\mathrm{ini})$ with $r_\mathrm{ini}$ satisfying $r(t, r_\mathrm{ini})< 0$.

Figure \ref{fig:j_insideout} shows the result of a calculation with $\cs=0.42 \kmps$ and the initial specific angular momentum profile with $j_0=1.8\times10^{-2} \kmps \pc$, which is roughly consistent with recent measurements of specific angular momentum in two Class 0 sources and a candidate of the first hydrostatic core \citep[][]{Pineda:2019aa}. The expected angular momentum profile at the age of 8$\times 10^{4}$ yrs, shown in Figure \ref{fig:j_insideout}, matches the measured profile including that of the $j$-increase regime at radii larger than 2900 au. The central stellar mass at the discussed age is $\sim$1.5$\Msun$, which is comparable to the protostellar mass of L1489 IRS. Note that the assumed $\cs$ corresponds to a temperature of $\sim50$ K, whereas a typical temperature of dense cores is $\sim$10 K \citep{Benson:1989aa}. These results suggest that the initial core of L1489 IRS was about five times more massive or denser than the one in a hydrostatic equilibrium. Such a massive initial dense core is actually expected for L1489 IRS because the infalling velocity slower than the freefall velocity suggests that an additional force supports infalling material against the gravity, as discussed in the next section. Note that computed specific angular momentum profiles basically do not change except for the timescale even if slow infall is considered.

Another analytical model proposed by \cite{Takahashi:2016aa} was also compared to the case of L1489 IRS. The model calculates the collapse of a super-critical Bonnor--Ebert sphere assuming that the entire dense core collapses simultaneously without following the inside-out collapse. Gas pressure is considered in the model in addition to the gravity:
\begin{align}
    \frac{dv_r}{dt} = \frac{GM}{f r_\mathrm{ini} r} - \frac{GM}{r^2}.
\end{align}
$f(r_\mathrm{ini})$ is proportional to the ratio between the gravitational potential and the thermal energy density at $r_\mathrm{ini}$, $-\frac{GM}{r_\mathrm{ini} \beta c_\mathrm{s}^2}$, where $\beta(r_\mathrm{ini}) \equiv \frac{\partial \ln \rho}{\partial \ln r} = \frac{\partial \ln \rho_\mathrm{ini}}{\partial \ln r_\mathrm{ini}}$ assuming that $\frac{\partial \ln \rho}{\partial \ln r}$ is constant in time. From consideration similar to that for the inside-out collapse model, the relation between the location and time is derived as follows:
\begin{align}
    t & = - \sqrt{\frac{r_\mathrm{ini}^3}{2GM}} \int_{1}^{x} \frac{dx'}{\sqrt{f^{-1}\ln x' + x'^{-1} - 1}}.
    \label{eq:t_x_T16}
\end{align}

In the calculation, we adopt a typical core radius $R_\mathrm{core}=10000$ au, the central density $n_\mathrm{c}=2.3 \times 10^{5}~\mathrm{cm^{-3}}$ and $c_\mathrm{s}=0.2 \kmps$ for a Bonnor--Ebert sphere and set $f=3$ to have the Bonnor--Ebert sphere unstable and explain the protostellar mass of L1489 IRS.

Figure \ref{fig:j_T16} compares the observations and the calculation result assuming $j_0=9\times 10^{-3}\kmps \pc$. The angular momentum profile at 7.6$\times 10^{4}$ yrs after the collapse starts roughly matches the measured profile including that of the $j$-increase regime at $\sim$2900--5000 au. The central stellar mass was estimated to be $\sim$1.6$\Msun$, which is consistent with that of L1489 IRS. The age of protostar was calculated as $t_\mathrm{sys} - t_\mathrm{prop}$, where $t_\mathrm{sys}$ is the time after the collapse starts and $ t_\mathrm{prop}$ is the epoch of protostar formation \citep[see][for more detail]{Takahashi:2016aa}. In the calculations, the protostellar age was estimated to be $\sim2.8 \times 10^4$ yrs.

Our model calculations yield a protostellar age of $\sim$(3--8)$\times$10$^4$ yrs. The typical lifetime of Class I sources has been estimated to be $\sim$(0.9--5)$\times 10^{5}$ yrs from the relative number of sources in different evolutionary stages \citep{Wilking:1989aa, Kenyon:1990aa, Evans:2009aa, Dunham:2015aa, Kristensen:2018aa}. Thus, the minimum value of the typical lifetime of Class I sources is comparable to, or $\sim$3 times larger than, the protostellar ages calculated in our models. Note that the lifetime derived from the previous observations mentioned above represents median lifetime or half-life of a sample of Class I protostars, and true lifetime has some diversity \citep{Evans:2009aa, Kristensen:2018aa}. In fact, the lifetime of individual dense cores depends on the core density and can vary by order of magnitude \citep{Kirk:2005aa}. The protostellar ages estimated in our calculations can also be longer by a factor of $\sim$2--3 if the infall velocity is slower than the freefall velocity, as suggested in Section \ref{subsec:ana_modeling}, although current calculations assume the infall timescale is determined by the freefall time.

It should be noted that the analytical calculations adopted here were simplified, and did not consider the possible effects of magnetic field. Non-ideal MHD simulations of disk formation showed that the radial profile of the specific angular momentum in the collapsing dense core could have a shallow slope at radii of $\sim$100--6000 au when ambipolar diffusion was efficient in the simulations \citep{Zhao:2018aa}. This shallow profile from the simulations could be similar to the green-dashed curve at 1000--5000 au in Figure \ref{fig:j_T16}. However, the simulations stopped when the stellar mass was less than $0.1\Msun$ and the disk radius was $\sim$20 au, which are over ten times smaller compared with those of L1489 IRS. Thus, it is difficult to make a direct comparison. Further simulations covering the later evolutionary stages or various initial conditions to produce larger stellar masses and disk radii are required to understand the observational results including the effects of magnetic field.
\subsection{Stellar-Mass Feeding Zone}
Comparisons between simple analytical models and the observations demonstrate that the observed profile of the specific angular momentum can be explained by the gravitational collapse of dense cores with relatively greater specific angular momenta as compared with the value measured by \cite{Goodman:1993aa}. Based on these models, we estimate the radius of the SMFZ. The size of the SMFZ can be estimated by considering from which radius of the initial core the infalling material showing the $j$-constant regime comes, because material having a specific angular momentum smaller than that of the $j$-constant regime would have already accreted into the disk or central protostar. In the case of the model based on the inside-out collapse, shown in Figure \ref{fig:j_insideout}, the radius of the SMFZ was estimated to be $\sim$4000 au by extrapolating the $j$-constant profile to the initial distribution of the specific angular momentum shown by the black line. Similarly, in the case of the collapsing super-critical Bonnor--Ebert sphere, shown in Figure \ref{fig:j_T16}, the radius of the SMFZ was estimated to be $\sim$6000 au.

The estimated radius of the SMFZ may not be the final value for L1489 IRS because infall from the envelope still continues. Regardless, we expect little material to accrete to the disk and the protostar because L1489 IRS is close to the end of Class I stage. The disk radius of L1489 IRS, $\sim$600 au, which is comparable to the maximum disk radius of single Class \II~sources in the Taurus star forming region \citep[$\sim$640 au;][]{Guilloteau:2014aa, Simon:2017aa}, also suggests that little material accretes to the disk and the protostar in L1489 IRS. Even if the disk radius of L1489 IRS increases from $\sim$600 au to 640 au, the size of SMFZ increases only 10\%, which is less significant.

It would be interesting to estimate sizes of SMFZs for Class \II~objects in the post-accretion phase because these sizes should be the final ones. It is, however, difficult to estimate them for lack of information constraining specific angular momentum distributions in initial dense cores for Class \II~objects without infalling envelopes. We should stress that Class I objects embedded in infalling envelope, such as L1489 IRS, allow us to measure the specific angular momentum in the outermost envelopes, constraining specific angular momentum distributions in initial dense cores, as shown in Figures \ref{fig:j_insideout} and \ref{fig:j_T16}.

It is interesting to note that the estimated radii of the SMFZ are smaller than that of typical dense cores. In the model based on a super-critical Bonnor--Ebert sphere, $R_\mathrm{core}=$10000 au was assumed and the estimated radius of the SMFZ was 60\% of $R_\mathrm{core}$. The total mass enclosed within 10000 au was $\sim$3.4$\Msun$, while the protostellar mass was $\sim$1.6$\Msun$, suggesting that $\sim$50\% of the total mass in the initial dense core forms a star. This value is close to the star formation efficiency suggested from comparisons between the CMF and IMF, $\sim$40\% \citep{Andre:2014aa, Konyves:2015aa}, and our result based on the model calculations suggest that only a limited area of the dense core forms the protostar(s) within the timescale of the Class 0 and I stages. The physical mechanism limiting the area of dense cores to form stars has to be studied further in the future.

%
\subsection{Regulated Infalling Velocity \label{subsec:slow_infall}}
Our modeling, presented in Section \ref{subsec:ana_modeling}, suggests that the infalling velocity in the envelope of L1489 IRS is slower than the freefall velocity yielded from the stellar mass of $1.6 \Msun$ by a factor of $\sim$2.5. This implies that some force supports infalling material against the gravitational force of the central protostar.

Magnetic field possibly supplies an additional force against gravity. Although neither the strength nor morphology of magnetic field in L1489 IRS is known, we can roughly estimate the possible strength of the magnetic field based on assumptions in the same way as discussed in \cite{Aso:2015aa}. We consider infalling motion along the envelope equatorial plane in cylindrical coordinates. The cylindrical radius $R$ is taken to be the radius in the equatorial plane. It is assumed that the magnetic field is perpendicular to the equatorial plane, and pinched toward the center. Such pinched magnetic fields showing the hourglass shape have been found in infalling envelopes mostly around Class 0 protostars \citep[e.g.,][]{Girart:2006aa}. Although L1489 IRS is a more evolved source, we adopted it here since very few measurements of magnetic fields exist at disk to envelope scales for Class I protostars. The equation of motion can be described as follows under a symmetric condition, where $\partial/\partial \theta=0 $ and $\partial/\partial z = 0$ \citep[][]{Aso:2015aa}:
\begin{align}
\rho(R) v_r (R) \frac{dv_r}{dR} & = -\frac{GM_\ast \rho(R)}{R^2} - \frac{1}{2\mu_0}\frac{dB^2}{dR} + \frac{B(R)^2}{\mu_0 R_\mathrm{curv}(R)}, \label{eq:eq_of_motion}
\end{align}
where $\Rcurv$ is the curvature radius of the magnetic field.
We rewrite the equation in a more convenient form assuming the radial dependence of $B$ as $B_0\left(\frac{R}{R_0}\right)^{-q}$ (see Appendix \ref{sec:app_eq_bfield} for more detail):
\begin{align}
B(R)& = \left[\frac{GM_\ast\rho(R)}{R^2} (1- \alpha^2) \mu_0
\left\{ \frac{q}{R} + \frac{1}{\Rcurv(R)}
\right\}^{-1}
\right]^{\frac{1}{2}}.
\label{eq:B_r}
\end{align}
For an order estimate, $\Rcurv=R$ was assumed because the magnetic field morphology is unknown. Then, from the best model parameters of $\alpha=0.4$ and $\rho_0=1.4 \times 10^{-18}\ \mathrm{g\ cm^{-3}}$, $B$ was estimated to be $\sim$0.25 mG at a radius of 1000 au from Equation (\ref{eq:B_r}), assuming $B\propto \rho^{2/3}$, which is a case of spherical symmetric collapse.

Zeeman observations on cloud and core scales have reported that the strength of the magnetic fields range from $\sim$1 $\mu$G to 1 mG. \citep{Crutcher:2010aa}. Observations in polarized continuum emission, which traces magnetic field directions, also estimated the magnetic-field strength on 100--1000 au scales to be an order of magnitude of mG through the Chandrasekhar--Fermi (CF) method \citep{Chandrasekhar:1953aa} or the alternative method proposed by \cite{Koch:2012aa} \citep{Girart:2006aa,Rao:2009aa,Stephens:2013aa}. Recent Zeeman observations of the Class \II~source TW Hya using ALMA in CN lines have reported non-detection of Zeeman spliting in the CN line from the protoplanetary disk around TW Hya, providing an upper limit of $|B_\mathrm{z}|<0.8$ mG \citep{Vlemmings:2019aa}. The magnetic-field strength estimated for L1489 IRS is within the range of observationally suggested values. Hence, the slow infall in L1489 IRS could be explained by the magnetic field if the magnetic-field strength in L1489 IRS is similar to that in other sources.

Similarly, a slow infalling velocity with $\alpha=0.3$ has been reported in the Class I protostar TMC-1A \citep{Aso:2015aa}. \cite{Aso:2015aa} estimated the strength of the magnetic field in TMC-A to be $\sim$2 mG at 200 au, which corresponds to 0.4 mG at 1000 au if extrapolated using the relation of $B\propto\rho^{2/3}$. Although L1489 IRS has larger stellar mass than TMC-1A (0.68$\Msun$), the values of $\alpha$ and the estimated strengths of the magnetic fields in both sources are similar. It is also reported that the infalling velocity in the Class 0 protostar L1527 IRS and the Class I protostar L1551 IRS 5 are slower than the expected freefall velocity by a factor of $\sim$2 and $\sim$3, respectively \citep{Ohashi:2014aa, Chou:2014aa}. Interestingly, all of these four sources suggest similar $\alpha$ values of 0.3--0.5 regardless of their different evolutionary stages. Slower infalling velocity could be common in the protostellar phase.
%
\section{Summary \label{sec:summary}}
We have conducted mapping observations covering a $\sim\ang[angle-symbol-over-decimal]{;2;} \times \ang[angle-symbol-over-decimal]{;2;}$ region around the protostar L1489 IRS using ACA and IRAM-30m in the \ce{C^18O} 2--1 emission in order to kinematically investigate a zone feeding mass into the central star, the ``stellar-mass feeding zone'' (SMFZ). The main results and conclusions are summarized as follows:
\begin{enumerate}
    \item We detected intensity peaks of both the 1.3 mm continuum and \ce{C^18O} emission at the protostellar position, which traced an envelope associated with the protostar. The \ce{C^18O} emission shows a velocity gradient at $r\gtrsim$1000 au in almost the same direction as that of the velocity gradient due to its disk rotation. The 1.3 mm continuum emission exhibits an intensity peak at the east side of the protostar, which is associated with a starless core. The \ce{C^18O} emission shows the second intensity peak at the northeast side of the protostar, suggesting another associated component.
    \item The peak velocity was measured as a function of radius at radii of $\sim$1000--5000 au on the southwest side of the protostar with the PV diagram cut along the major axis of the disk rotation to characterize the observed velocity gradient. The measured peak velocity decreases with radius inside 2900 au, suggesting the differential rotation of the envelope, but is constant or slightly increasing outside 2900 au.
    \item The measured peak velocity is considered to trace the rotational velocity of the envelope based on comparison of the disk-and-envelope models to the observations. The model with slow infall, wherein the infalling velocity is slower than the freefall velocity by a factor of 2.5 and the specific angular momentum is constant, best reproduced the observed PV diagram and the radial profile of the measured peak velocity within 2900 au. The model also suggested that the measured peak velocity was almost comparable to the expected rotational velocity of the envelope with a constant specific angular momentum. In order to explain the increase of the peak velocity outside 2900 au in radius, the specific angular momentum has to increase outside the radius of 2900 au in the model.
    \item We calculated the radial profile of the specific angular momentum at $\sim$300--5000 au from the peak velocity measured in current and previous works. The specific angular momentum profile consists of three parts: a Keplerian disk inside $\sim$600 au, $j$-constant regime at $\sim$600--2900 au, and $j$-increase regime outside $\sim$2900 au. We compared it to analytic models of collapsing dense cores having a finite angular momentum, assuming that infalling materials conserve their specific angular momenta. The measured specific angular momentum profile is explained by the models. The analytic models suggest that the initial core more massive or denser than a singular isothermal sphere or a critical Bonnor--Ebert sphere is preferred to explain the protostellar mass of L1489 IRS of 1.6$\Msun$.
    \item Based on comparison of the radial profiles of the specific angular momentum derived from the observations and the analytical models, we estimated the size of the SMFZ, a zone where material forming the protostar resides in the initial dense core, to be $\sim$4000--6000 au in radius in L1489 IRS. These radii are significantly smaller than typical radii of dense cores of $\sim$10000--20000, suggesting that only a part of the initial core feeds the protostar.
    \item The infalling velocity slower than the freefall velocity by a factor of $\sim$2.5 can be explained by the magnetic field with a strength of $B\sim$0.25 mG, assuming the morphology of the magnetic field. The estimated strength of magnetic field is approximately comparable to that measured in molecular clouds and protostellar envelopes with previous observations.
\end{enumerate}

\acknowledgments
This paper used the following ALMA data: ADS/JAO.ALMA \#2019.1.01063.S. ALMA is a partnership of ESO (representing its member states), NSF (USA), and NINS (Japan), together with NRC (Canada), MOST and ASIAA (Taiwan), and KASI (Republic of Korea), in cooperation with the Republic of Chile. The Joint ALMA Observatory is operated by ESO, AUI/NRAO, and NAOJ. We thank all ALMA staff for conducting the observations. This work is based on observations carried out under project number 136-18 with the IRAM-30m telescope. IRAM is supported by INSU/CNRS (France), MPG (Germany) and IGN (Spain). We would like to thank the IRAM staff for their support during the campaigns, and the NIKA2 collaboration for enabling tools to reduce the data. We also thank St\'{e}phane Guilloteau for the fruitful discussions on the merging of the ACA and IRAM-30m data. We are grateful to Kengo Tomida, Doug Johnstone, Yuri Aikawa, Patrick M.~Kock, Ya-Wen Tang, and Masayuki Yamaguchi for the productive discussions. J.S.~thanks Shigehisa Takakuwa for the financial support for his trip to the IRAM-30m telescope to make observations. A.M.~is supported by the European Research Council (ERC) under the European Union Horizon 2020 research and innovation program (MagneticYSOs project, grant agreement No. 679937). J.S.~is supported by Academia Sinica Institute of Astronomy and Astrophysics.

\facility{ALMA, IRAM-30m}
\software{CASA \citep{McMullin:2007aa}, GILDAS (\url{http://www.iram.fr/IRAMFR/GILDAS}), RADMC-3D \citep{Dullemond:2012aa}, Numpy \citep{Oliphant:2006aa,van-der-Walt:2011aa}, Scipy \citep{Jones:2001aa}, Astropy \citep{Astropy-Collaboration:2013aa,Astropy-Collaboration:2018aa}, Matplotlib \citep{Hunter:2007aa}}

\newpage
\restartappendixnumbering
\appendix
%
\section{Supplemental Model \label{sec:app_model}}
Three supplemental models are provided to explore further possibilities explaining the increase of the peak velocity outside the radius of 2900 au.

\subsection{Disk-and-envelope model with modified $v_r$ \label{subsec:app_vrmodel}}
Disk-and-envelope models with modified $v_r$, where infalling velocity increases outside the break radius, were compared with the observations to examine whether such models could explain the peak velocity increasing outside 2900 au. Two models following Equation (\ref{eq:rho_model}) to (\ref{eq:vphi_model}) except for the radial velocity distribution were constructed with the parameters in Table \ref{tab:model_params} and $\alpha=0.4$. One model, referred to as $v_r$-model 1, has the following radial velocity distribution:
\begin{align}
    v_r(r, \theta) =
    \begin{cases}
    -\alpha \left(\frac{GM_\ast}{r} \right)^{0.5} \left(1 + \frac{\cos \theta}{\cos \theta_0} \right)^{0.5} & (r \leq \rbreak) \\
    -v_{r,\mathrm{break}} \left(\frac{r}{\rbreak} \right)^{p_\mathrm{inf}} & (r > \rbreak)
    \end{cases},
    \label{eq:vr_sup01}
\end{align}
where $v_{r,\mathrm{break}}=0.4 \kmps$ and $p_\mathrm{inf}=1.2$ so that the radial velocity increases with radius outside the break radius and reaches the freefall velocity at the edge of the envelope. The other model, $v_r$-model 2, has the radial velocity distribution of
\begin{align}
    v_r(r, \theta) =
    \begin{cases}
    -\alpha \left(\frac{GM_\ast}{r} \right)^{0.5} \left(1 + \frac{\cos \theta}{\cos \theta_0} \right)^{0.5} & (r \leq \rbreak) \\
    - \left(\frac{GM_\ast}{r} \right)^{0.5} \left(1 + \frac{\cos \theta}{\cos \theta_0} \right)^{0.5} & (r > \rbreak)
    \end{cases},
    \label{eq:vr_sup02}
\end{align}
so that the radial velocity is slower than the freefall velocity inside the break radius but is the freefall velocity outside.

\begin{figure*}[htpb]
\centering
\includegraphics[width=2\columnwidth]{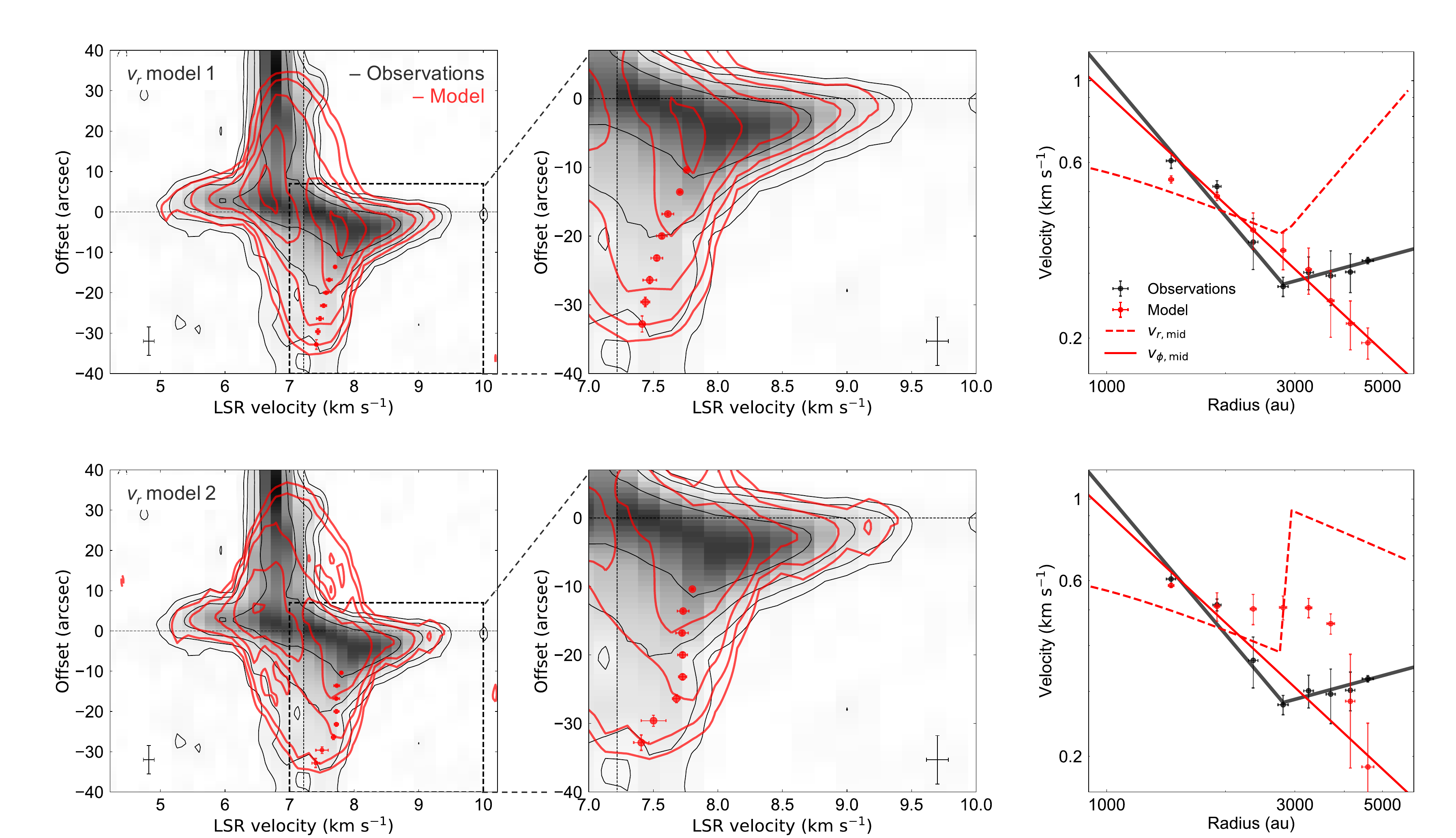}
\caption{Same as Figure \ref{fig:pv_obsvsmodels} except for the two supplemental models, which have modified radial velocity distributions. \label{fig:app_model_vr}}
\end{figure*}

Figure \ref{fig:app_model_vr} shows the model results. The solid, dashed lines in the third column in Figure \ref{fig:app_model_vr} show the given radial velocity for the models. Although both models have radial velocities increasing outside the break radius, the observed peak velocity cannot be reproduced.

\subsection{Disk-and-flow model \label{subsec:app_flowmodel}}
\begin{figure*}[htpb]
\centering
\includegraphics[width=2\columnwidth]{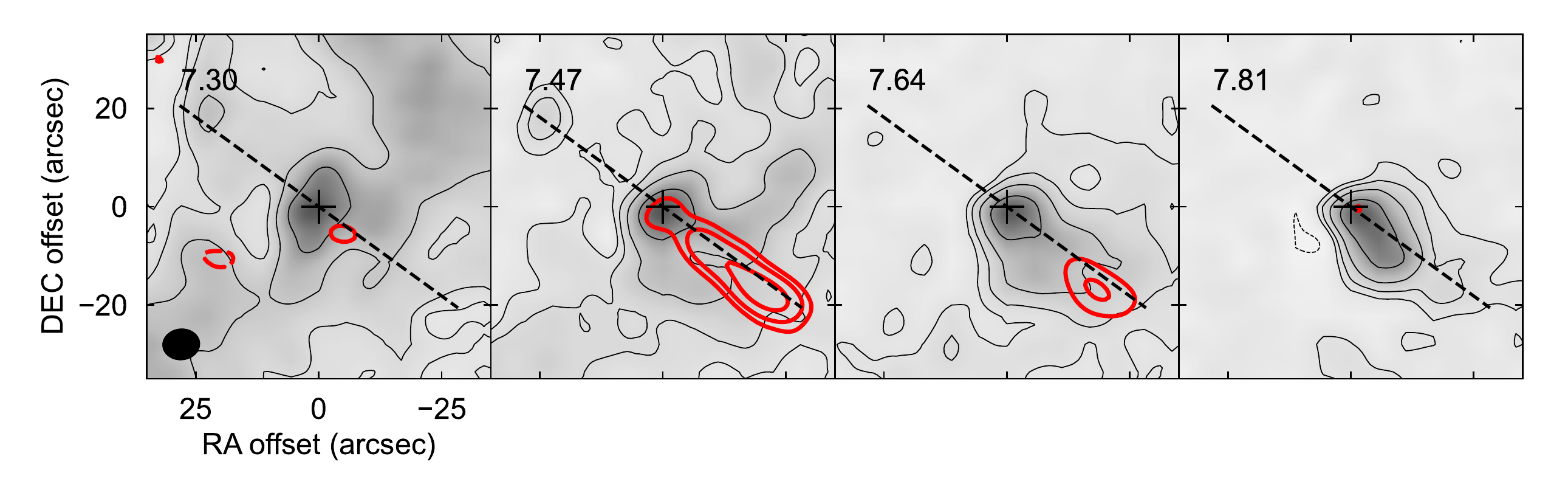}
\caption{Velocity channel maps of the disk-and-flow model (red contours) overlaid on the channel maps of the observations (background color and black contours). Contour levels and symbols are the same as those in Figure \ref{fig:chan_c18o}. Dashed lines show the direction of the PV cut (P.A.$=$54$^\circ$). \label{fig:app_model_flow_channel}}
\end{figure*}

\begin{figure*}[htpb]
\centering
\includegraphics[width=1.8\columnwidth]{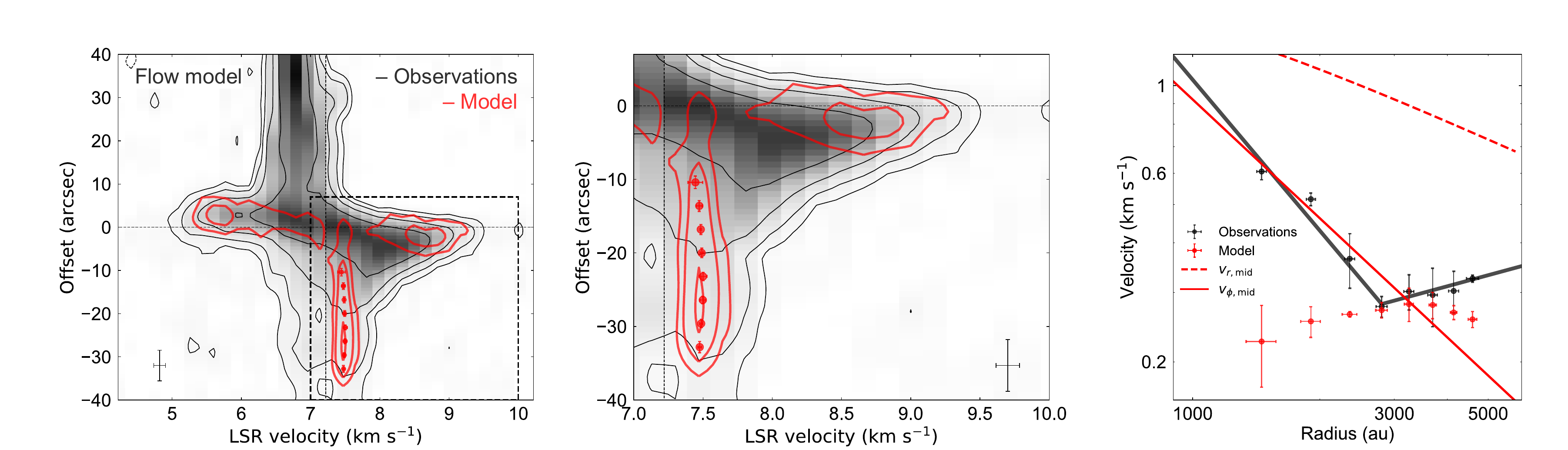}
\caption{Same as Figure \ref{fig:pv_obsvsmodels} except for the disk-and-flow model. \label{fig:app_model_flow_pv}}
\end{figure*}

A disk-and-flow model was built to investigate whether flow-like infall such as those reported by \cite{Yen:2014aa} and \cite{Pineda:2020aa} can explain the observed velocity structure outside the break radius or not. We modified the envelope structure of the disk-and-envelope model described in Section \ref{subsec:ana_modeling} to build a disk-and-flow model having a flow falling from certain directions. The direction of the flow is determined by the parameters of $\theta_0$ and $\phi_0$, which are the initial location of the infalling material in $\theta$ and $\phi$ at an infinite radius, respectively. We adopted $\theta_0=95 \pm 5^\circ$, where $\theta=90^\circ$ is the disk mid-plane, so that the direction of the flow is along the direction of the disk major axis (i.e., the direction of the PV cut). A case with $\phi_0$ of $326 \pm 2^\circ$ was compared with the observations. Other parameters describing the envelope structures are the same as those of the base disk-and-envelope model summarized in Table \ref{tab:model_params}, except for the volume density $\rho_0$, which was fixed at $4 \times 10^{-17} \gperccm$. The emission from the flow component of the disk-and-flow model is presented in Figure \ref{fig:app_model_flow_channel}, showing the trajectory and velocity of the flow projected on the plane-of-sky.

The PV diagram of the disk-and-flow model cut along the disk major axis shown in Figure \ref{fig:app_model_flow_pv} demonstrates that the model does not reproduce the observed PV diagram zoomed in the middle panel of Figure \ref{fig:app_model_flow_pv}. The radial profile of the peak velocity measured in the disk-and-flow model, presented in the third column of Figure \ref{fig:app_model_flow_pv}, shows that the measured peak velocity slightly increases with radius at $\sim$1400--4000 au regardless of the decrease of $v_r$ and $v_\phi$. This is because the infalling flow is more parallel to the plane-of-sky at inner radii, resulting in smaller line-of-sight velocity at smaller radii. Although the measured peak velocity of the model cannot explain the observed profile of the peak velocity inside the break radius, it is roughly consistent with the observed profile outside the break radius.

It should be noted that the direction of the flow has to be very specific to explain the observed velocity structure outside the break radius. The adopted initial angles of  $(\theta_0,\phi_0)=(95\pm5^\circ,326\pm2^\circ)$ were analytically estimated so that the flow reproduces the observed velocity structure outside the break radius best. Although we have investigated different sets of the initial angles and flow widths, the increase of the peak velocities outside the break radius was not reproduced better than the current model.

\section{\ce{N_2H^+} $J=$1--0 \label{sec:app_n2hp}}
\begin{figure}[thb]
\includegraphics[viewport=40 0 802 595, width=\columnwidth]{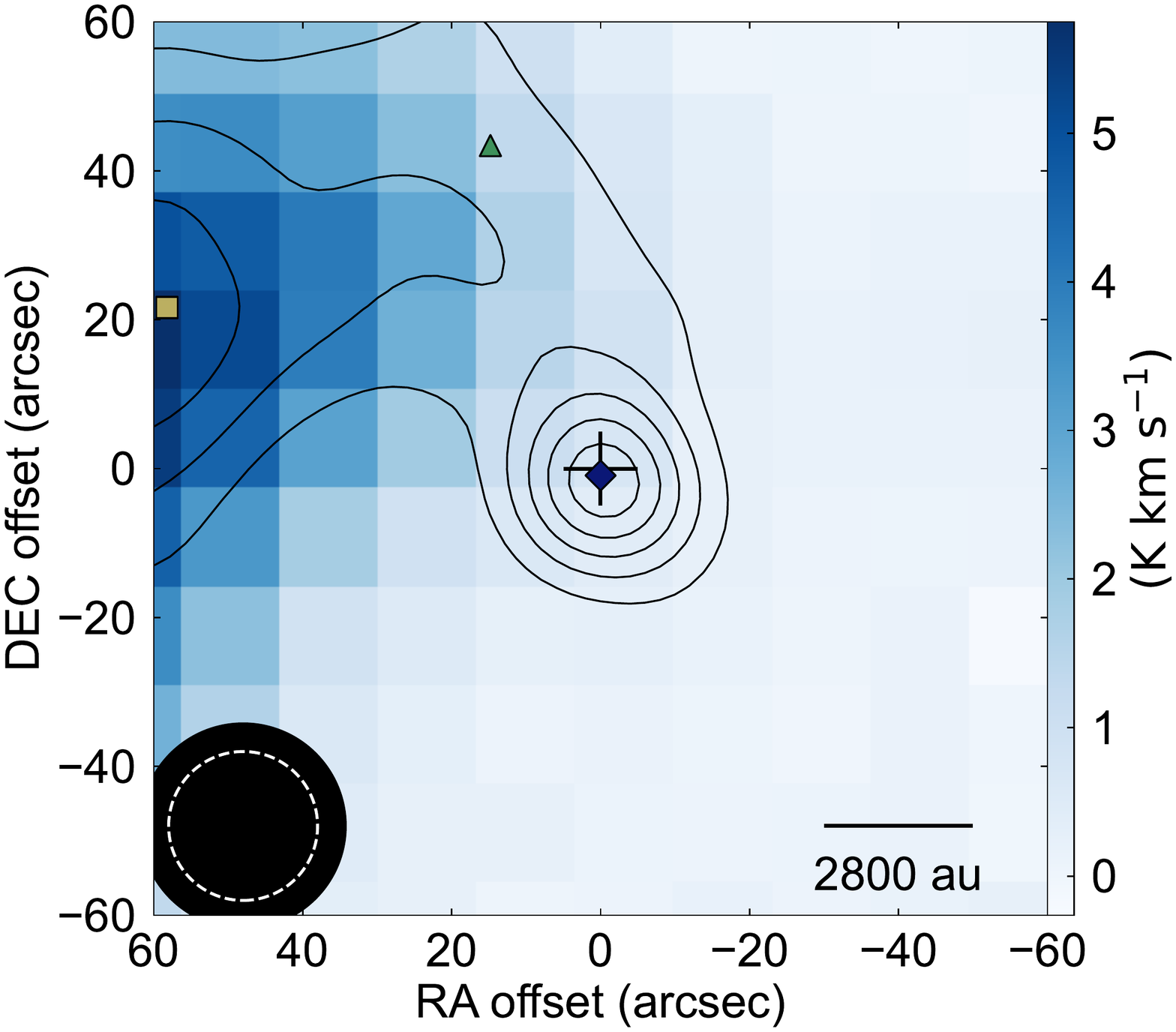}
\caption{The moment 0 map of the \ce{N_2H^+} 1--0 (background color) and the 1.3 mm continuum map (contours). Contour levels start at 3$\sigma$ and increases by steps of 2$\sigma$, where $1\sigma$ corresponds to 5 $\mjpbm$. The dashed circle and the filled ellipse in the bottom-left corner denote the beam size for the 1.3 mm continuum map and the \ce{N_2H^+} map, respectively. The central cross denotes the protostellar position. Colored markers show positions where column densities are derived. \label{fig:mom0_n2hp}}
\end{figure}

We present the data of \ce{N_2H^+} 1--0 and a brief discussion about the spatial distributions of the \ce{N_2H^+} 1--0 and \ce{C^18O} 2--1 emissions. Figure \ref{fig:mom0_n2hp} shows the moment 0 map of the \ce{N_2H^+} emission with the 1.3 mm continuum map. The \ce{N_2H^+} emission was integrated over a velocity range of $-1.8$--13.9$\kmps$ to include all hyperfine components of the emission. The \ce{N_2H^+} emission shows a peak at the position of the 1.3 mm continuum second peak, and is likely associated with the starless core at the eastern side of the protostar. The overall distribution of the \ce{N_2H^+} emission is consistent with previous observations \citep{Caselli:2002aa}. The spatial distributions of the \ce{C^18O} and \ce{N_2H^+} emissions are apparently anti-correlated (the primary and secondary peaks of the \ce{C^18O} 2--1 emission are denoted by a blue diamond and a green triangle, respectively, in Figure \ref{fig:mom0_n2hp}).

To investigate the anti-correlation of these two molecular emissions quantitatively, the abundances of \ce{C^18O} and \ce{N_2H^+} molecules were estimated from their column densities and the \ce{H2} column densities. The measurements were made at three different points shown in Figure \ref{fig:mom0_n2hp}: the \ce{C^18O} primary and secondary peaks, and the \ce{N_2H^+} peak. We used \ce{C^18O} and continuum images smoothed with the beam of the \ce{N_2H^+} map for calculations.
%

Dust column densities are derived from the following equation assuming an optically thin emission:
\begin{align}
    \Sigma_\mathrm{dust} = \frac{I_\nu}{B_\nu(T_\mathrm{dust})\kappa_\nu},
    \label{eq:Sigma_dust}
\end{align}
where $I_\nu$ is the intensity, $B_\nu$ is the Plank function, $T$ is the dust temperature, and $\kappa_\nu$ is the dust-mass opacity. For the dust temperature,  it was assumed that $T_\mathrm{dust}=20$ K at the \ce{C^18O} primary peak position and $T_\mathrm{dust}=10$ K at the other two positions. We applied $\kappa_\nu=0.90\ \mathrm{g\ cm^{-2}}$ derived by \cite{Ossenkopf:1994aa}, the gas-to-dust mass ratio of 100, and the mean molecular weight $\mu=2.8$ \citep{Kauffmann:2008aa} to derive the dust column density and convert it to the column density of \ce{H_2} gas.

The method described in \cite{Caselli:2002ac} was adopted to calculate the column densities of \ce{C^18O} and \ce{N_2H^+} molecules. For the \ce{C^18O} emission, $T_\mathrm{ex}=20$ K at the \ce{C^18O} primary peak position and $T_\mathrm{ex}=10$ K at the other two positions were assumed. For the \ce{N_2H^+} emission, we performed fitting to its hyperfine components with the python package \textit{PySpecKit} \citep[][]{Ginsburg:2011aa} and derived the \ce{N_2H^+} column density from the total opacity $\tau_\mathrm{tot}$, the intrinsic velocity dispersion $\sigma_v$, and the excitation temperature $\tex$. Table \ref{tab:fitres_hyperfine} lists the results of the fitting to the hyperfine components of the \ce{N_2H^+} emission at three positions. Molecular abundances of \ce{C^18O} and \ce{N_2H^+} were derived from the ratio of the column density of each molecule to that of the \ce{H_2} gas. The derived column densities and molecular abundances are summarized in Table \ref{tab:column_density}.

The derived \ce{N_2H^+} abundances are similar within a factor of 2 among the three measurement points, while those of \ce{C^18O} are different between the \ce{C^18O} primary/secondary peaks and the \ce{N_2H^+} peak by a factor of $\sim$3--4. Because of the difference of \ce{C^18O} abundances, the abundance ratios of $X_\mathrm{\ce{C^18O}}/X_\mathrm{\ce{N_2H^+}}$ at the \ce{C^18O} primary/secondary peaks are $\sim$4--6 times larger than that at the \ce{N_2H^+} peak.

\ce{C^18O} abundances at the \ce{C^18O} primary/secondary peaks are comparable to the molecular cloud values in the Taurus region \citep{Frerking:1982aa}, and to the values measured in other Class I objects \citep[][]{Jorgensen:2002aa}. A possible reason for the lower \ce{C^18O} abundance at the \ce{N_2H^+} peak would be that \ce{CO} molecules including \ce{C^18O} freeze out onto dust grains, which is often seen in starless cores \citep[e.g.,][]{Bergin:2002aa, Tafalla:2004aa}. The \ce{N_2H^+} emission is likely associated with the L1489 starless core \citep{Motte:2001aa, Wu:2019aa}, supporting our idea of a \ce{CO} freeze-out at the \ce{N_2H^+} peak. \cite{Aikawa:2015aa} shows a relation between \ce{CO} abundance and the \ce{H_2} gas temperature based on a calculation of chemical network. Their calculation suggests that \ce{CO} abundance can be reduced by an order of magnitude due to a \ce{CO} freeze-out. Thus, the lower abundance nearby the center of the starless core could be reproduced by \ce{CO} freeze-out. Molecular abundance of \ce{C^18O} at the \ce{C^18O} second peak is similar to that measured at protostellar position. However, the \ce{C^18O} second peak is at the edge of the starless core \citep{Motte:2001aa,Wu:2019aa} and the volume density is expected to be lower than at the center of the starless core, resulting in longer timescale for \ce{CO} freeze-out. Photodesorption would also be more effective at lower density, suppressing the \ce{CO} depletion at the \ce{C^18O} second peak. Hence, the anti-correlation between the \ce{N_2H^+} and \ce{C^18O} emissions would be the \ce{CO} freeze-out at the starless core.
%
\begin{table*}[th!]
    \centering
    \caption{Results of the fitting to hyperfine components of the \ce{N_2H^+} 1--0 emissions}
    \begin{tabular*}{2\columnwidth}{@{\extracolsep{\fill}}lcccccc}
    \hline
    \hline
    Position & RA & Dec. & $\tex$ & $\tau_\mathrm{tot}$ & $v_\mathrm{cent}$ & $\sigma_v$ \\
     & (J2000) & (J2000) & (K) & & ($\kmps$) & ($\kmps$) \\
    \hline
     \ce{C^18O} primary peak & 04:04:43.07 & +26:18:55.28 & 3.1$\pm0.1$ & 4.6$\pm1.8$ & $6.96\pm0.02$ & $0.23 \pm 0.01 $ \\
     \ce{C^18O} secondary peak & 04:04:44.17 & +26:19:39.59 & 4.1$\pm0.2$ & $3.7\pm0.6$ & $6.758 \pm 0.003$ & $0.148 \pm 0.003$ \\
     \ce{N_2H^+} peak & 04:04:47.40 & +26:19:17.85 & $6.58\pm0.04$ & $8.9\pm 0.2$ & $6.7779\pm 0.0005$ & $0.1115\pm 0.0005$ \\
    \hline
    \end{tabular*}
\label{tab:fitres_hyperfine}
\end{table*}
%
\begin{table*}[th!]
    \centering
    \caption{Measured column densities and molecular abundances}
    \begin{tabular*}{2\columnwidth}{@{\extracolsep{\fill}}lcccccc}
    \hline
    \hline
    Position & $N_{\ce{H_2}}$ & $N_{\ce{C^18O}}$ & $N_{\ce{N_2H^+}}$ & $X_{\ce{C^18O}}$ & $X_{\ce{N_2H^+}}$ & $X_{\ce{C^18O}}/X_{\ce{N_2H^+}}$ \\
     & ($\times 10^{21}\ \mathrm{cm^{-2}}$) & ($\times 10^{15}\ \mathrm{cm^{-2}}$) & ($\times 10^{12}\ \mathrm{cm^{-2}}$) & ($\times 10^{-7}$) & ($\times 10^{-10}$) & ($\times 10^{2}$)\\
    \hline
     \ce{C^18O} primary peak & 8.7$\pm$0.2 & 1.211$\pm$0.009 & 4.3$\pm$1.7 & 1.39$\pm$0.03 & 4.9$\pm$2.0 & 2.8$\pm$1.1  \\
     \ce{C^18O} secondary peak & 12.4$\pm$0.6 & 1.53$\pm$0.01 & 3.2$\pm$0.5 & 1.23$\pm$0.06 & 2.6$\pm$0.4 & 4.8$\pm$0.8 \\
     \ce{N_2H^+} peak & 24.1$\pm$0.6 & 0.92$\pm$0.01 & 11.6$\pm$0.3 & 0.38$\pm$0.01 & 4.8$\pm$0.2 & 0.79$\pm$0.04 \\
    \hline
    \end{tabular*}
\label{tab:column_density}
\end{table*}
%
\section{Derivation of Equation (14) \label{sec:app_eq_bfield}}
Here, we present the derivation of Equation (\ref{eq:B_r}). The radial component of the equation of motion is as follows \citep{Aso:2015aa}:
\begin{align}
\rho(R) v_r (R) \frac{dv_r}{dR} & = -\frac{GM_\ast \rho(R)}{R^2} - \frac{1}{2\mu_0}\frac{dB^2}{dR} + \frac{B(R)^2}{\mu_0 R_\mathrm{curv}(R)}, \label{eqap:eq_of_motion} \\
 & = \Fg + \Fb, \nonumber
\end{align}
where
\begin{align}
\Fg &= -\frac{GM_\ast \rho(R)}{R^2},\label{eqap:fg} \\
\Fb &= \Fmp + \Fmt \nonumber\\
& = - \frac{1}{2\mu_0}\frac{dB^2}{dR} + \frac{B(R)^2}{\mu_0 R_\mathrm{curv}(R)}. \label{eqap:fb}
\end{align}
Here, $\Fmp$ is the magnetic pressure and $\Fmt$ is the magnetic tension. $\Fg$ is always negative from Equation (\ref{eqap:fg}).

Now, we assume $v_r (R)$ to be $v_r(R) = \alpha \vff (R)$, where $\vff (r)=-\sqrt{2G M_\ast/r}$ is the freefall velocity, and $\alpha$ is constant. Note that this assumption means that the potential is negative at any radius and zero at an infinite radius, i.e., $|\Fg| > |\Fb|$. The derivative of $v_r$ is, 
\begin{align}
\frac{dv_r(R)}{dR} &= \alpha \frac{d\vff}{dR} \nonumber \\
& = \alpha \frac{1}{2} \sqrt{\frac{2GM_\ast}{R^3}}.
\end{align}
Therefore, the left-hand side of Equation (\ref{eqap:eq_of_motion}) can be written as,
\begin{align}
\rho(R) v_r (R) \frac{dv_r}{dR} &= -\rho(R) \alpha^2 \frac{GM_\ast}{R^2}\nonumber\\
&=\alpha^2 \Fg. \label{eqap:2}
\end{align}
Thus, from Equation (\ref{eqap:eq_of_motion}) and (\ref{eqap:2}),
\begin{align}
\alpha^2 = \left( 1 + \frac{\Fb}{\Fg}
\right), \label{eqap:alpha2} \\
\therefore \alpha = \sqrt{\left( 1 + \frac{\Fb}{\Fg}
\right)}.
\end{align}

Assuming $B=B_0\left(\frac{R}{R_0}\right)^{-q}$, the term of $\Fmp$ can be written as follows:
\begin{align}
\Fmp &= -\frac{1}{2\mu_0}\frac{dB^2}{dR}\\
&= \frac{q}{\mu_0} \frac{B_0^2}{R_0} \left(\frac{R}{R_0} \right)^{-2q-1}\nonumber\\
& = \frac{1}{\mu_0} B_0^2 \left(\frac{R}{R_0} \right)^{-2q} \frac{q}{R}\nonumber\\
& =  \frac{1}{\mu_0} B(R)^2 \frac{q}{R}.
\end{align}
Therefore, from Equation (\ref{eqap:fb}), we obtain
\begin{align}
\Fb = \frac{B(R)^2}{\mu_0} \left\{\frac{q}{R} + \frac{1}{\Rcurv(R)} \right\}.
\end{align}
From this equation and Equation (\ref{eqap:alpha2}), we obtain the magnetic field $B$ as a function of $R$ and $\alpha$:
\begin{align}
\Fb = \frac{B(R)^2}{{\mu_0}}& \left\{ \frac{q}{R} + \frac{1}{\Rcurv(R)} \right\} = \Fg (\alpha^2 -1),\nonumber
\end{align}
\begin{align}
\therefore B(R)& = \left[\Fg (\alpha^2 -1) \mu_0
\left\{ \frac{q}{R} + \frac{1}{\Rcurv(R)}
\right\}^{-1}
\right]^{\frac{1}{2}}.
\end{align}
Note that $\Fb/\Fg$ has no R-dependence according to Equation (\ref{eqap:alpha2}) because $\alpha$ is assumed to be constant here.
%

\bibliography{reference}{}
\bibliographystyle{aasjournal}



\end{document}